\documentclass{article}
\usepackage[final]{neurips_2025}
\usepackage[utf8]{inputenc} 
\usepackage[T1]{fontenc}    
\usepackage{hyperref}       
\usepackage{url}            
\usepackage{booktabs}       
\usepackage{amsfonts}       
\usepackage{nicefrac}       
\usepackage{microtype}      
\usepackage{xcolor}         
\usepackage{tcolorbox}
\usepackage{fancyvrb}
\usepackage{multicol}
\usepackage{multirow}       
\usepackage{amsmath}
\usepackage{listings}
\usepackage{courier}
\usepackage{enumitem}
\usepackage{tabularx}
\usepackage{wrapfig}

\usepackage{enumitem}
\setlist[itemize]{leftmargin=2.5em, itemsep=0.2em}
\setlist[enumerate]{leftmargin=2.5em, itemsep=0.2em}

\title{CodeAssistBench (CAB): Dataset \& Benchmarking for Multi-turn Chat-Based Code Assistance}

\author{%
  Myeongsoo Kim\\
  AWS AI Labs\\
  \texttt{mysoo@amazon.com}
  \And
  Shweta Garg \\
  AWS AI Labs \\
  \texttt{shwegarg@amazon.com} \\
  \AND
  Baishakhi Ray \\
  AWS AI Labs \\
  \texttt{rabaisha@amazon.com} \\
  \And
  Varun Kumar \\
  AWS AI Labs \\
  \texttt{kuvrun@amazon.com} \\
  \And
  Anoop Deoras \\
  AWS AI Labs \\
  \texttt{adeoras@amazon.com} \\
}

\begin{document}
\maketitle
\begin{abstract}
Programming assistants powered by large language models have improved dramatically, yet existing benchmarks still evaluate them in narrow code-generation settings. Recent efforts such as InfiBench and StackEval rely on Stack Overflow questions and remain limited to single-turn interactions, manually curated data, and isolated snippets rather than full project environments. We introduce CodeAssistBench (CAB), the first benchmark for evaluating multi-turn, project-grounded programming assistance at scale. CAB automatically constructs datasets from GitHub issues tagged as questions, using an LLM-driven pipeline that filters noise, extracts runnable contexts, builds executable containers, and verifies environment correctness. This enables continuous, automated expansion across diverse repositories without manual intervention. Using CAB, we create a testbed of 3,286 real-world issues across 214 repositories, spanning seven languages. Evaluating state-of-the-art models reveals a substantial gap: while models achieve 70–83\% accuracy on Stack Overflow–style questions, they solve only 7.22–16.49\% of CAB issues from post-training-cutoff repositories. These results highlight a fundamental challenge: current LLMs struggle to provide assistance in realistic, project-specific contexts despite strong performance on traditional Q\&A benchmarks. CAB provides a scalable, reproducible framework for advancing research in multi-turn, codebase-grounded programming agents. The benchmark and pipeline are fully automated and publicly available at https://github.com/amazon-science/CodeAssistBench/.
\end{abstract}

\section{Introduction}
\label{sec:introduction}

Large language models (LLMs) are increasingly integrated into modern programming workflows, supporting tasks ranging from code generation to debugging and repository navigation. While benchmarks have progressed from isolated synthesis tasks (HumanEval~\cite{chen2021evaluating}, MBPP~\cite{austin2021program}) to repository-level maintenance settings (SWE-Bench~\cite{jimenez2023swe}, BigCodeBench~\cite{zhuo2024bigcodebench}), they still capture only a narrow slice of how developers seek assistance in practice. Recent multi-turn benchmarks such as ConvCodeWorld~\cite{han2025convcodeworld}, MINT~\cite{wang2023mint}, and TICODER~\cite{fakhoury2024ticoder} move toward conversational evaluation, yet they primarily emphasize code synthesis and assume stable, well-defined contexts.

However, real-world programming assistance extends far beyond writing code. A 2024 Stack Overflow survey of 34{,}168 developers reports that while 75.7\% seek AI for code generation, even higher percentages desire support for searching for answers (77.9\%), debugging and troubleshooting (77.3\%), and understanding unfamiliar codebases (73.6\%)~\cite{stackoverflow2024aisurvey}. These tasks require iterative clarification, environment-aware reasoning, and integration with project-specific details—capabilities that current benchmarks largely overlook.

Recent efforts such as InfiBench~\cite{li2024infibench} and StackEval~\cite{shah2024stackeval} incorporate Stack Overflow data to better reflect developer Q\&A scenarios, but they remain limited to single-turn interactions over isolated snippets and require substantial manual curation.

\begin{wrapfigure}{r}{0.32\textwidth}
  \centering
  \vspace{-10pt}
  \includegraphics[width=0.30\textwidth]{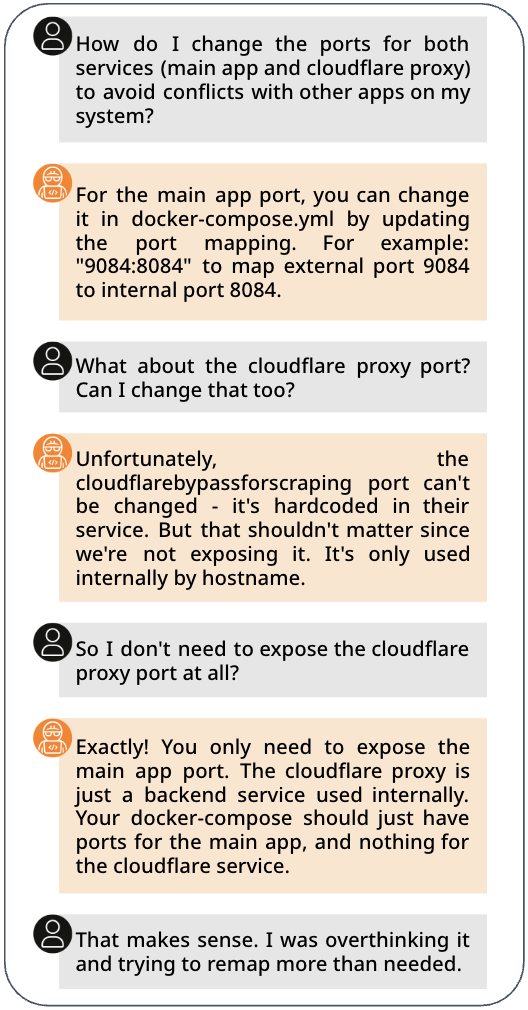}
  \caption{A port-mapping clarification from a real GitHub issue.}
  \label{fig:port-conflict}
  \vspace{-15pt}
\end{wrapfigure}

Real assistance is inherently interactive. Even seemingly simple questions often require the assistant to infer project topology, reference environment configuration, or reconcile ambiguous user descriptions. Figure~\ref{fig:port-conflict} illustrates a real GitHub issue in which a user asks about container port mapping. Solving it requires the assistant to (1) interpret the repository's network structure, (2) explain that the proxy port is internal and hard-coded, and (3) assure the user that no additional mapping is needed. Each answer influences the next question, defining a trajectory of reasoning and communication. Evaluating such scenarios cannot rely solely on execution oracles or single-shot correctness; it requires tracking how well the assistant guides the user across turns with clarity, efficiency, and accuracy.

To address these limitations, we introduce \textbf{CodeAssistBench (CAB)}, the first benchmark designed to evaluate multi-turn, project-grounded programming assistance at scale. CAB automatically constructs datasets from GitHub issues tagged as questions using an LLM-driven pipeline that filters noisy issues, extracts runnable contexts, builds containerized environments, and verifies their correctness. This automation enables continuous, reproducible benchmark expansion without manual curation and supports dataset variants drawn from repositories created after major model training cutoffs, ensuring persistent difficulty over time.

CAB consists of 3,286 real-world programming questions across 214 repositories in seven languages, paired with an execution-backed, chat-driven evaluation pipeline. A simulated user provides contextual feedback, a maintainer agent interacts with the full codebase in a containerized environment, and an automated judge assesses conversation quality using extracted satisfaction conditions. This framework captures the full complexity of real-world developer–assistant interactions: ambiguous problem descriptions, evolving hypotheses, and multi-turn negotiation of understanding.

Our evaluation of leading LLMs reveals a striking capability gap. While recent models achieve 70--83\% accuracy on Stack Overflow-style questions, they solve only 16.49\% of CAB issues from post-training-cutoff repositories. These results highlight a fundamental challenge: despite impressive performance on traditional Q\&A benchmarks, LLMs struggle to provide reliable assistance in realistic, multi-turn, project-specific contexts.

CAB provides a scalable and reproducible foundation for advancing research in multi-turn programming assistance. By coupling fully automated dataset generation with environment-grounded conversational evaluation, it offers a new lens through which to measure and improve the real-world capabilities of LLM-based developer tools.

\section{Related Work}
\label{related_work}

Although programming assistance spans diverse developer needs, existing benchmarks only capture narrow aspects of real-world workflows.

\textbf{Code Generation Benchmarks.}
Traditional evaluations of programming assistants focus on generating functionally correct code from well-specified prompts. HumanEval~\cite{chen2021evaluating}, MBPP~\cite{austin2021program}, and CodeContests~\cite{li2022competition} assess single-shot code synthesis, while more advanced benchmarks such as SWE-Bench~\cite{jimenez2023swe} and SWE-PolyBench~\cite{rashid2025swepolybenchmultilanguagebenchmarkrepository} use real GitHub issues to evaluate repository-level maintenance tasks. These benchmarks provide robust execution-based correctness metrics but remain predominantly single-turn and do not capture the iterative, conversational nature of real developer interactions.

\textbf{Multi-turn Programming and Conversational Benchmarks.}
Several benchmarks incorporate iterative interactions. ConvCodeWorld~\cite{han2025convcodeworld}, MINT~\cite{wang2023mint}, and TICODER~\cite{fakhoury2024ticoder} simulate developer--assistant dialogues but typically assume stable, fully specified environments and remain centered on code synthesis tasks. More general conversational evaluations—such as MT-Bench~\cite{zheng2023judging}, AlpacaEval~\cite{dubois2023alpacafarm}, and BotChat~\cite{duan-etal-2024-botchat}—measure dialogue quality but are not targeted at technical problem-solving in programming contexts. In contrast, we evaluate multi-turn, task-grounded interactions where assistance depends on project state, ambiguous context, and evolving user needs.

\textbf{Programming Q\&A Benchmarks.}
StackEval~\cite{shah2024stackeval} and InfiBench~\cite{li2024infibench} use Stack Overflow data to better reflect practical developer questions and emphasize explanatory reasoning. However, they remain limited to single-turn settings, require substantial manual curation, and lack integration with executable project environments. These constraints prevent them from capturing the iterative reasoning, context reconstruction, and environment-aware troubleshooting that dominate real-world programming assistance.

Overall, prior work evaluates code generation, multi-turn dialogue, or Q\&A reasoning in isolation. \textbf{CAB is the first benchmark to combine multi-turn conversational evaluation with full-project, environment-grounded contexts generated entirely automatically from real GitHub workflows.}

\section{CAB: CodeAssistBench}
\label{sec:CAB}

\begin{figure}[t]
  \centering
  \includegraphics[width=\linewidth]{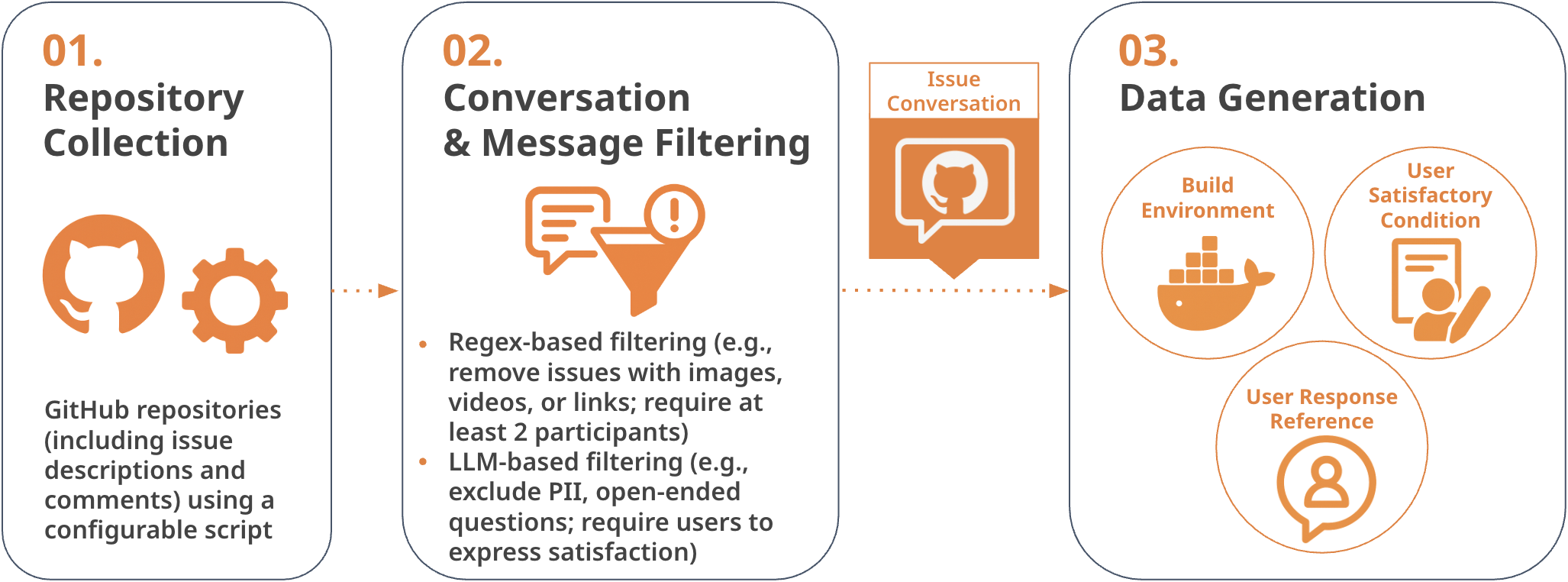}
  \vspace{-10pt}
  \caption{CAB’s automated data generation pipeline. It collects relevant GitHub repositories, filters issue conversations, and produces structured assistance scenarios with build environments, satisfaction conditions, and reference user responses.}
  \vspace{-15pt}
  \label{fig:cab-datageneration}
\end{figure}

CodeAssistBench (CAB) is a fully automated benchmark framework for evaluating multi-turn, project-grounded programming assistance. CAB converts real GitHub issues into structured, executable assistance scenarios and evaluates LLMs through simulated multi-agent interactions. The framework has two components: (1) an automated dataset construction pipeline and (2) an environment-grounded conversational evaluation system.

\subsection{Dataset Generation Pipeline}
\label{sec:dataset-generation}

CAB transforms raw GitHub issues into structured assistance tasks through three stages: 
(1) repository collection, 
(2) issue filtering, and 
(3) data preparation. 
An overview is shown in Figure~\ref{fig:cab-datageneration}.

\subsubsection{Repository Collection}
\label{sec:repo-collection}

We begin by collecting a diverse set of GitHub repositories that represent real-world programming challenges across multiple languages. Let $\mathcal{R}_{GH}$ denote the set of all public GitHub repositories. We extract a filtered subset:
\begin{equation}
R = \left\{ r \in \mathcal{R}_{GH} \mid s(r) > S_{\min},\; t(r) > t_0,\; \ell(r) \in \mathcal{L} \right\}
\end{equation}
where $s(r)$ is the star count of repository $r$ with threshold $S_{\min}$ (e.g., 10), $t(r)$ is its creation date with cutoff $t_0$ (e.g., \texttt{2024-11-01}), and $\ell(r)$ is $r$'s SPDX license, restricted to permissive licenses compatible with research use.

To prioritize repositories with active developer support communities, we define a community score $\mathrm{CS}(r) = Q(r) + H(r)$, where $Q(r)$ and $H(r)$ represents the number of closed issues labeled with ``question'' and ``help wanted'' respectively. We then select the top-$N$ repositories ranked by $\mathrm{CS}(r)$ (breaking ties by star count) to form our repository set $R_N$, where $N$ is a user-defined parameter.

\subsubsection{Issue Filtering}
\label{sec:conv-msg-filter}

For each repository $r \in R_N$, let $I_r$ be the set of issues associated with it. We fetch only successfully closed issues using the GitHub Issues API. Each issue is represented as:
\begin{equation}
i = (\mathrm{title}(i),\,\mathrm{body}(i),\,\mathrm{messages}(i)), \quad \mathrm{messages}(i) = (m_{i,1}, m_{i,2}, \dots)
\end{equation}

We apply two filtering rules - implemented using regular expressions - to retain high-quality, interactive conversations: (1) requiring at least two distinct participants to ensure genuine question-answering dynamics; and (2) removing issues containing media content (e.g., URLs, images, videos) to focus on text-based programming assistance.

To ensure issue relevance and message quality, we apply two LLM-based filtering steps using structured prompts. At the issue level, we assess resolution status, specificity, clarity, and safety via a seven question prompt (see Listing~\ref{lst:issue-prompt}), asking the model to answer each of them with a binary Yes/No. Issues are retained only if they satisfy criteria such as being clearly resolved, technically specific, free of sensitive content (e.g., personal information), and reproducible. At the message level, we construct the full conversation and prompt the model to identify comments that provide no support-related value (e.g., "+1", "Thanks", "Bump"). Comments are preserved unless explicitly flagged for removal using strict exclusion rules (see Listing~\ref{lst:comment-prompt}). 

After filtering, we restructure each issue as a sequence of turns, where a turn represents a complete user-maintainer interaction:
\begin{equation}
\mathrm{turn}_{i,k} = (m_{i,k}^{\mathrm{author}}, m_{i,k}^{\mathrm{maintainer}})
\end{equation}
To construct turns, we group consecutive messages from the same role (author or maintainer) into a single logical message, then pair each author segment with the subsequent maintainer segment. Unpaired trailing messages (e.g., a final ``thanks'' from the author) are discarded.

The resulting filtered issue set $I'_r$ contains issues with this turn-based structure:
\begin{equation}
i = (\mathrm{title}(i),\,\mathrm{body}(i),\,\{\mathrm{turn}_{i,1}, \mathrm{turn}_{i,2}, \ldots\})
\end{equation}

\subsubsection{Data Preparation}
\label{sec:data-generation}

For each filtered issue, we prepare three essential components: build environment generation, user satisfaction condition extraction, and user response reference generation.

\textbf{Build Environment Generation.} For issues requiring environment-specific testing, we automatically generate Docker configurations by analyzing repository artifacts using Sonnet 3.7~\cite{claude37sonnet} with a structured prompt (Appendix~\ref{sec:dockerfile}) that identifies the commit $\mathit{sha}_i$ closest to the issue creation timestamp, clones the repository at that commit, and extracts key artifacts (README content, Dockerfiles, GitHub workflows, file structure). It then generates and tests candidate build scripts until finding a successful configuration $e_i$.

\textbf{Satisfaction Condition Extraction.} We use Sonnet 3.7 with a structured prompt (provided in Appendix~\ref{sec:satcond}) to identify explicit criteria that indicate when an issue is successfully resolved. The model analyzes the full conversation, focusing on the original question and subsequent clarifications, to extract concrete conditions that would satisfy the user's needs. These form the set $s_i$ = $\{s_{i,1}, \dots, s_{i,K}\}$ of satisfaction conditions that serve as objective evaluation criteria.

\textbf{User Response Reference Generation.} To enable realistic simulation of user follow-up behavior, we construct a BM25 index over historical maintainer-user message pairs (excluding data from the current issue) and retrieve the top-$N$ most similar maintainer responses with their corresponding user replies. These form the reference set $u_i$ = $\{u_{i,1}, \dots, u_{i,M}\}$ that guides the simulated user's feedback.

\textbf{Result Aggregation.} The resulting dataset entry for each issue is a tuple $d_i = \bigl(r_i,\;i,\;\mathit{sha}_i,\;e_i,\;s_i,\;u_i\bigr)$. For each issue \(i \in \bigcup_r I_r'\), where \(I_r'\) denotes the filtered issues for repository \(r\), we create such entries to form our complete dataset.

\subsection{Evaluation Framework and Implementation}
\label{sec:evaluation-framework}

\begin{figure}[ht]
  \centering
  \includegraphics[width=0.8\linewidth]{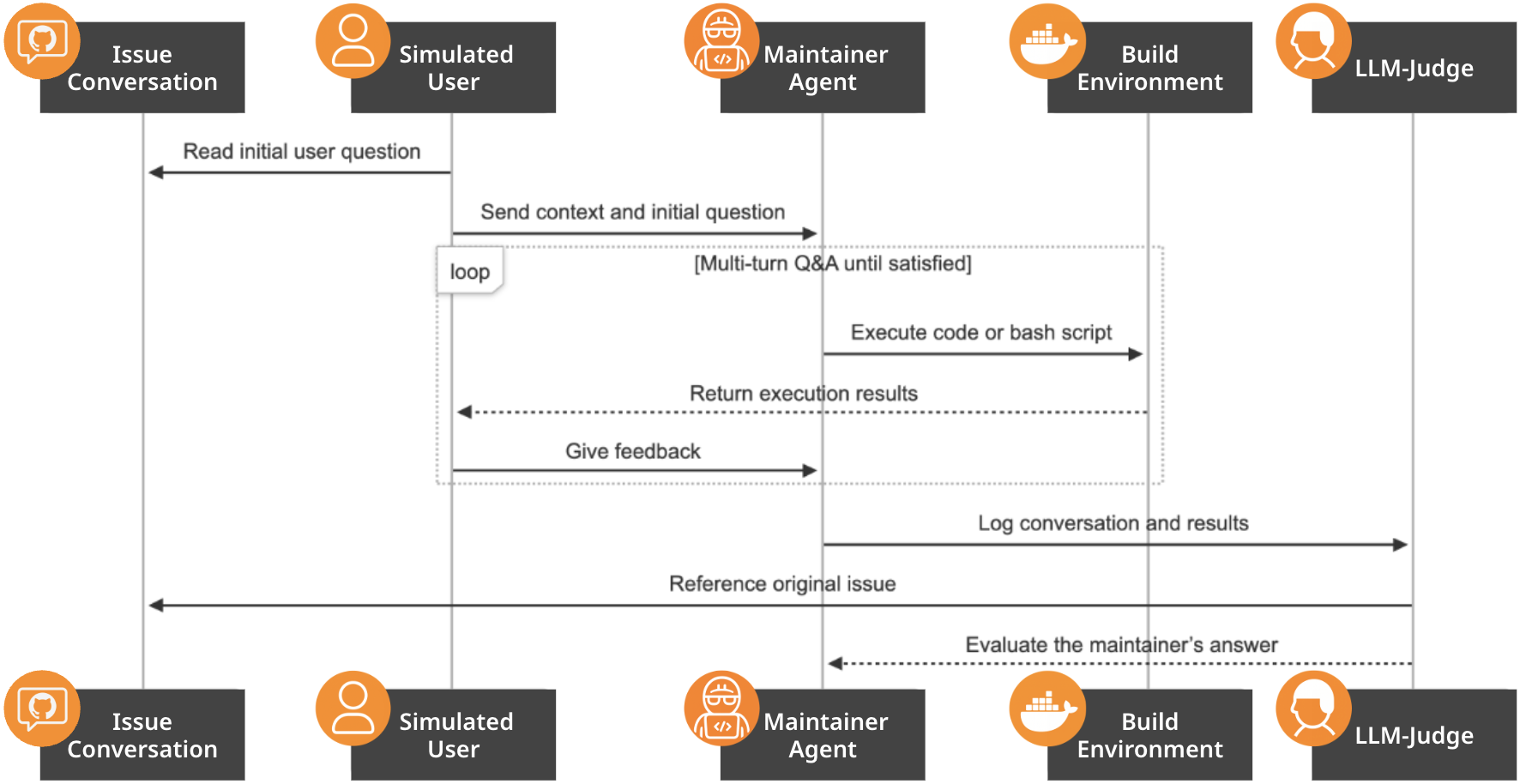}
\caption{\textbf{CAB evaluation pipeline.} A simulated user chats with the Maintainer Agent, which can run code in an optional build sandbox; the interaction continues until the user is satisfied or reaches the maximum turn limit. Once the dialogue ends, an LLM judge grades the exchange against satisfaction conditions extracted from the original GitHub issue. This pipeline enables realistic assessment of programming assistance in context-rich, project-specific scenarios.}
  \label{fig:cab-evaluation-appendix}
\end{figure}

CAB's evaluation framework simulates realistic programming assistance interactions through a multi-agent system with three distinct roles: a User, a Maintainer, and a Judge (see Figure~\ref{fig:cab-evaluation-appendix}).

\textbf{User Agent.}
The user agent initiates the conversation with a programming question from a GitHub issue and provides follow-up responses. It presents the initial question with context, evaluates model responses against satisfaction conditions, provides realistic follow-ups or clarifications, and signals when a solution resolves the issue. The user agent observes execution results but does not directly interact with the environment.

\textbf{Maintainer Agent.}
The maintainer agent represents the LLM being evaluated. Given the user's question and contextual information, the model must analyze the problem within the provided codebase context, execute commands in the containerized environment when necessary, generate helpful responses, and adapt its approach based on user feedback.

\textbf{Judge Agent.}
After the conversation concludes, an automated judge evaluates the interaction quality based on: (1) technical correctness - whether the proposed solution correctly addresses the underlying issue; (2) satisfaction completeness - whether all extracted satisfaction conditions are met; and (3) interaction quality - whether the conversation was appropriately concise and helpful. For issues with Docker environments, execution success is a hard requirement - a technically sound explanation is insufficient if the implementation fails in practice. The evaluation is triggered when either the user expresses satisfaction or the conversation reaches a maximum of 10 turns (configurable). This approach provides a comprehensive assessment of an LLM's ability to provide effective programming assistance in realistic scenarios.

CAB is implemented as a fully automated pipeline that continuously generates new benchmark datasets as GitHub repositories evolve. Our implementation integrates with GitHub's REST API to collect repositories and issues with configurable parameters, automatically generates and validates Docker environments to ensure reproducibility, and orchestrates the multi-agent conversation flow while recording detailed interaction logs. This enables CAB to generate continuously evolving benchmarks that remain challenging as models improve, particularly through dataset variants including repositories created after model training cutoff dates.

\section{Experimental Results and Analysis}
\label{sec:results}
We evaluate CAB along three dimensions: (1) dataset generation outcomes, (2) model performance in multi-turn debugging, and (3) human studies validating judge reliability and satisfaction-condition quality. Hardware details are provided in Appendix~\ref{sec:requirements}.

\subsection{Statistics of Dataset Generation Pipeline}
\label{sec:dataset-generation-eval}

\begin{table}[t]
\centering
\caption{Summary of issue filtering, Docker requirement detection, build outcomes, and final issue retention across programming languages and repository cohorts in our dataset generation pipeline. The 56-issue difference between filtered issues (3,342) and final retained issues (3,286) corresponds to Docker-required issues where environment builds failed (294 Docker-required $-$ 238 successful builds $=$ 56 excluded).}
\resizebox{0.75\linewidth}{!}{
\begin{tabular}{l|r|r|r|r|r|r|r|r}
\toprule
\textbf{Metric} & \textbf{Total} & \textbf{Python} & \textbf{Java} & \textbf{C++} & \textbf{C\#} & \textbf{JS} & \textbf{TS} & \textbf{C} \\
\midrule
\textbf{Filtered Issues}        & 3,342 & 660 & 543 & 518 & 545 & 445 & 443 & 188 \\
\quad All-Time                  & 3,033 & 500 & 535 & 487 & 520 & 406 & 417 & 168 \\
\quad Recent                    & 309   & 160 & 8   & 31  & 25  & 39  & 26  & 20  \\
\cmidrule(lr){1-9}
\textbf{Docker-Required}        & 294   & 90  & 58  & 53  & 30  & 34  & 12  & 17  \\
\quad All-Time                  & 252   & 64  & 58  & 48  & 28  & 30  & 9   & 15  \\
\quad Recent                    & 42    & 26  & 0   & 5   & 2   & 4   & 3   & 2   \\
\cmidrule(lr){1-9}
\textbf{Successful Docker Builds} & 238 & 77  & 52  & 28  & 21  & 33  & 10  & 17  \\
\quad All-Time                  & 197   & 52  & 52  & 23  & 19  & 29  & 7   & 15  \\
\quad Recent                    & 41    & 25  & 0   & 5   & 2   & 4   & 3   & 2   \\
\cmidrule(lr){1-9}
\textbf{Final Retained Issues} & 3,286 & 647 & 537 & 493 & 536 & 444 & 441 & 188 \\
\quad All-Time                  & 2,978 & 488 & 529 & 462 & 511 & 405 & 415 & 168 \\
\quad Recent                    & 308   & 159 & 8   & 31  & 25  & 39  & 26  & 20  \\
\bottomrule
\end{tabular}
}
\vspace{-5pt}
\label{tab:dataset_stats}
\end{table}

We applied our automated pipeline (Section~\ref{sec:dataset-generation}) to two repository cohorts: an \textbf{All-Time} set of 700 top-starred repositories (100 per language) without creation-date limits, and a \textbf{Recent} set of 3{,}500 repositories (500 per language) created after November~2024. After applying license and community filters, we retain 770 repositories for downstream processing. These produced 25{,}656 raw GitHub issues tagged with \texttt{question} or \texttt{help-wanted}. We apply a layered filtering process, regex heuristics followed by an LLM classifier, to remove low-signal or ambiguous issues. Issues requiring Docker environments are excluded if automated build validation fails.

Across both cohorts, we retained \textbf{3,286 multi-turn issues from 214 repositories}, including \textbf{238 fully validated Docker environments}. Many repositories yielded no qualifying multi-turn issues after filtering, resulting in 214 repositories contributing at least one benchmark instance. Notably, \textbf{Recent repositories achieved a 97.6\% build success rate} compared to 78.2\% for All-Time repositories (details in Appendix~\ref{sec:docker-build-details}). In total, our pipeline executed 44{,}628 Sonnet~3.7 calls end-to-end without any manual intervention. Issue-length distributions (Figure~\ref{fig:conv-length-distribution}) show that over half of all issues involve multi-turn threads, particularly in Python, C++, and TypeScript—highlighting the need for multi-step reasoning benchmarks such as CAB.

\subsection{Model Evaluation Results}
\label{sec:model-evaluation}

We benchmark six state-of-the-art LLMs as maintainer agents: ChatGPT 4.1 Mini, DeepSeek~R1, Llama~3.3~70B, Haiku~3.5, Sonnet~3.7, and Sonnet~3.7 (Think). Models interact with simulated users until all satisfaction conditions are met or the turn limit is reached.

\subsubsection{Evaluation Metrics}
We measure:
\begin{itemize}
\item \textbf{Correctness}: \textit{Correct}, \textit{Partially Correct}, or \textit{Incorrect}, determined by a judge LLM evaluating condition satisfaction, execution outcomes, and reference alignment.
\item \textbf{Verbosity}: \textit{Appropriate}, \textit{Verbose}, or \textit{Terse}, capturing explanation style and conciseness.
\end{itemize}
Judge prompts and scoring criteria are in Appendix~\ref{sec:judge-maintainer}.

\subsubsection{Sampling Strategy}
Running all 3,286 issues in multi-turn mode is computationally prohibitive. We therefore construct a balanced subset: per language, up to five Dockerfile issues plus ten issues per turn length bucket (1, 2, 3, 4, 5+). Underflow buckets reallocate to earlier ones. This yields \textbf{350 datasets} for All-Time repositories and \textbf{194 datasets} for Recent repositories.

\begin{table}[t!]
\centering
\caption{Correctness and average CAB turns. Rec.~=~Recent repositories; All~=~All-Time.}
\label{tab:combined-accuracy-corrected}
\setlength{\tabcolsep}{5pt}
\renewcommand{\arraystretch}{1.2}
\resizebox{\textwidth}{!}{%
\begin{tabular}{l|cc|cc|cc||cc|cc|cc}
\toprule
\multirow{2}{*}{\textbf{Model}} &
\multicolumn{6}{c||}{\textbf{Correctness (\%)}} &
\multicolumn{6}{c}{\textbf{Avg. CAB Turn}} \\
& \multicolumn{2}{c|}{Correct} & \multicolumn{2}{c|}{Partially Corr.} & \multicolumn{2}{c||}{Incorrect} 
& \multicolumn{2}{c|}{Correct} & \multicolumn{2}{c|}{Partially Corr.} & \multicolumn{2}{c}{Incorrect} \\
& Rec. & All & Rec. & All & Rec. & All 
& Rec. & All & Rec. & All & Rec. & All \\
\midrule
ChatGPT 4.1 Mini        & \textbf{16.49} & \textbf{29.14} & 30.41 & 36.86 & \textbf{53.09} & 34.00 & 2.94 & 2.35 & 3.66 & 3.33 & 5.70 & 4.28 \\
DeepSeek R1             & 11.34 & 27.14 & 33.51 & 40.29 & 55.15 & \textbf{32.57} & 2.82 & 2.24 & 3.18 & 2.72 & 4.50 & 4.28 \\
Llama 3.3 70B           &  9.33 & 13.58 & 25.91 & 40.75 & 64.77 & 45.66 & 3.22 & 2.68 & 4.06 & 3.49 & 4.67 & 4.50 \\
Haiku 3.5               &  7.22 & 16.86 & 30.93 & 43.43 & 61.86 & 39.71 & 3.86 & 2.73 & 3.97 & 3.81 & 6.76 & 5.63 \\
Sonnet 3.7              & 11.34 & 25.71 & 30.93 & 35.43 & 57.73 & 38.86 & 2.36 & 2.30 & 3.57 & 3.15 & 5.71 & 4.21 \\
Sonnet 3.7 (Think)      & 13.40 & 27.43 & 27.32 & 38.86 & 59.28 & 33.71 & 2.50 & 2.20 & 3.25 & 2.85 & 4.95 & 4.26 \\
\bottomrule
\end{tabular}%
}
\vspace{-15pt}
\end{table}

\subsubsection{Correctness Results} 

Table~\ref{tab:combined-accuracy-corrected} presents the performance metrics and conversation statistics for all evaluated models. We observe a consistent performance gap between recent and historical repositories. On recent repositories, correctness rates range from 7.22\% (Haiku 3.5) to 16.49\% (ChatGPT 4.1 Mini), while on all-time repositories, these rates increase substantially to 13.58-29.14\%. ChatGPT 4.1 Mini demonstrates superior performance across both datasets, achieving the highest correctness (16.49\% recent; 29.14\% all-time) and lowest error rates on recent repositories (53.09\%), while DeepSeek R1 shows the lowest error rate on all-time repositories (32.57\%).

To better understand the reason behind this recency gap, we investigate two possibilities: (1) model knowledge limitations, and (2) intrinsic properties of newer repositories. To isolate these effects, we construct a lightweight synthetic ablation using 50 automatically generated repositories (details in Appendix~\ref{sec:synthetic-ablation-appendix}). Surprisingly, Claude~3.7~Sonnet achieved 74\% correctness on these synthetic repositories compared to only 11.34\% on recent real repositories. The synthetic repositories contained richer documentation and version constraints within model training windows, suggesting that the recency gap is primarily driven by post–knowledge-cutoff framework changes rather than properties of AI-generated code. Full methodology and analysis appear in Appendix~\ref{sec:synthetic-ablation-appendix}.

\begin{figure*}[t]
\centering
\includegraphics[width=0.95\linewidth]{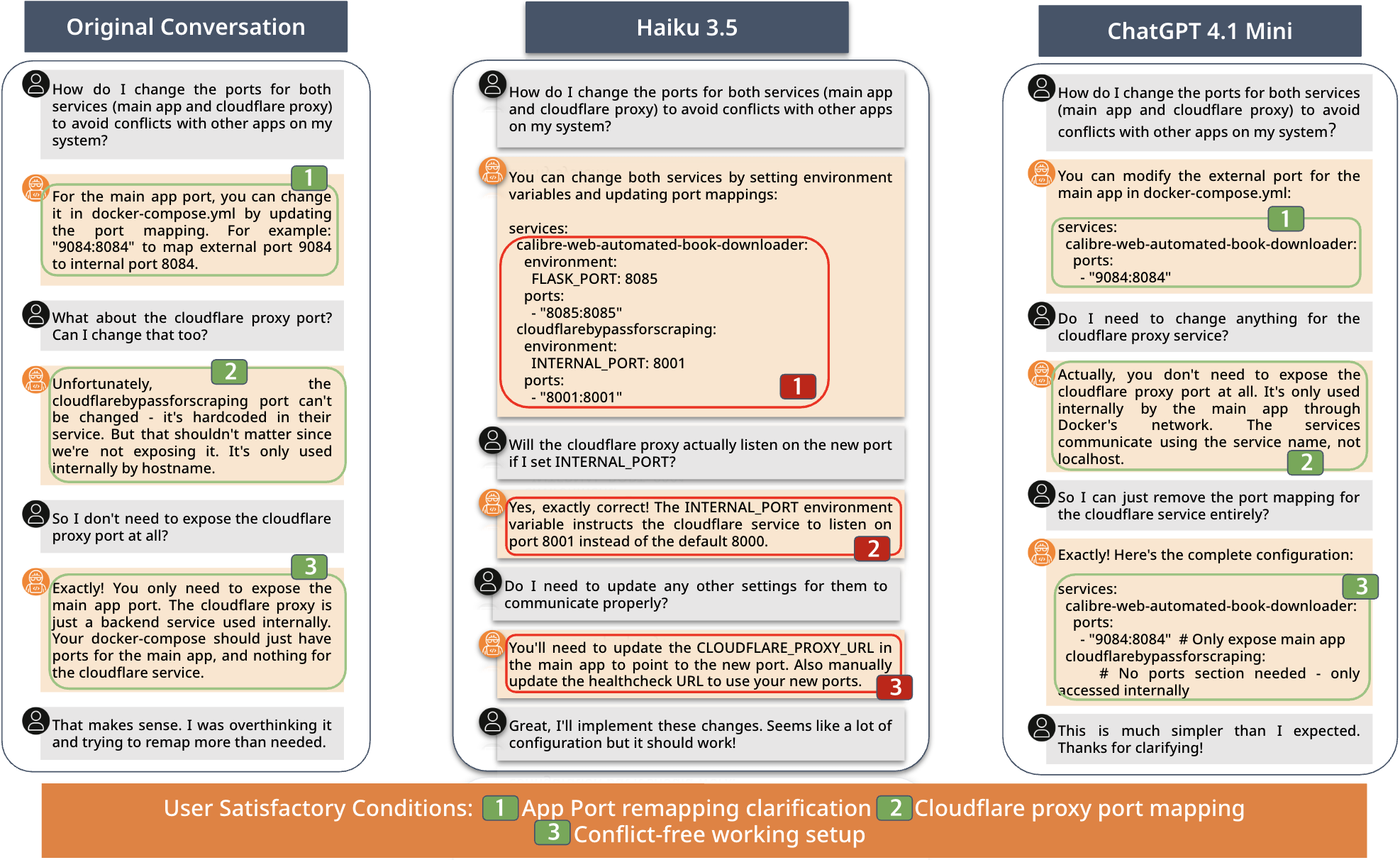}
\caption{Side-by-side comparison of model responses to a Docker port-remapping issue: Haiku 3.5's incomplete solution (middle) fails to address key requirements, while ChatGPT 4.1 Mini's successful response (right) satisfies all three user conditions (highlighted in colored boxes).} 
\label{fig:example}
\vspace{-10pt}
\end{figure*}

\subsubsection{Qualitative Example}
Figure~\ref{fig:example} illustrates a GitHub issue where the user asks how to remap ports for both the main application and the Cloudflare proxy to avoid conflicts with other apps. It presents two contrasting model \textit{stories}: Haiku 3.5 provides an incomplete response, either omitting clear proxy-port guidance or introducing incorrect configurations, while ChatGPT 4.1 Mini offers a successful solution - clearly explaining how to change the main app’s port, noting that the proxy port doesn’t need to be exposed, and providing a conflict-free \texttt{docker-compose.yml}. A simulated user agent drives each LLM through a multi-turn dialogue, asking follow-up questions until all satisfaction conditions are met or a predefined turn limit is reached. Where applicable, the agent also validates the recommended configuration using recorded build outcomes. Once the conversation concludes, an automated LLM judge assigns a final label - \textit{Correct}, \textit{Partially Correct}, or \textit{Incorrect} - based on comparisons with the original GitHub thread, execution results, and the predefined satisfaction conditions.

\subsubsection{Conversation-Length Analysis}

Table~\ref{tab:combined-accuracy-corrected} compares the average turn counts between original GitHub threads and CAB-generated dialogues. When models produce correct answers, CAB conversations are similar in length to real GitHub threads - typically spanning 2–3 turns. In contrast, incorrect cases tend to result in longer CAB dialogues, often by 1–2 additional turns. This suggests that when the model-as-maintainer provides vague or incomplete responses, the simulated user - like a real developer - asks follow-up questions for clarification.

\subsubsection{Verbosity Analysis}

Figure~\ref{fig:verbosity} (Appendix) summarizes verbosity classifications. Models tend toward over-explanation: verbose responses account for 40–60\% of outputs. Sonnet~3.7 (Think) shows the most balanced behavior, while Haiku~3.5 and Llama~3.3~70B are most verbose. Verbosity increases on recent repositories, likely due to higher uncertainty.

\subsubsection{Language-Specific Analysis}
\label{sec:language-analysis}

Figure~\ref{fig:languages} reports the percentage of \textit{Correct} responses - excluding partially correct answers - for each model across seven programming languages, split by recent and all-time GitHub repositories. The analysis reveals stark contrasts in language difficulty and model specialization. 

Statically typed languages such as C\#, C++, and Java remain particularly challenging, especially on recent issues. For instance, in the recent C\# dataset, most models achieve less than 13\% correctness. Similarly, correctness on recent C++ issues hovers below 15\% for all models except ChatGPT 4.1 Mini. These results suggest models often struggle with precision and strict type constraints in newer repositories.

By contrast, dynamically typed languages like JavaScript and Python show relatively stronger performance on the all-time dataset. ChatGPT 4.1 Mini and Sonnet 3.7 Think both reach 44\% correctness on JavaScript, while DeepSeek R1 and Sonnet 3.7 Think achieve over 30\% on Python and TypeScript. Nevertheless, performance on recent repositories remains low across the board. Even for JavaScript, the best model - Sonnet 3.7 Think - achieves only 15.4\% correctness. These results highlight the increased complexity or novelty of recent code issues.

Overall, Sonnet 3.7 Think (M6) consistently ranks among the top performers in JavaScript, TypeScript, and Python. ChatGPT 4.1 Mini (M1) also shows strength in all-time datasets for JavaScript and TypeScript. No model, however, maintains high correctness across both typed and dynamic languages, particularly in recent repositories.

\subsection{Human Evaluation Studies}
\label{sec:human-alignment}

We conducted two human evaluation studies to validate CAB's automated components: judge reliability and satisfaction condition quality.

\subsubsection{Judge Validation}

To assess the reliability of our automated LLM judge, two software engineers 
(3+ years experience) independently evaluated 310 model responses sampled 
across languages, difficulty levels, and issue types. The annotators achieved 
substantial inter-rater agreement (78.28\%, Cohen's $\kappa = 0.68$; see 
Table~\ref{tab:inter-rater} in Appendix~\ref{sec:expert-validation-appendix}), 
establishing a strong human baseline.

Comparing the automated judge to human annotators, the LLM judge achieved 
65.92\% average agreement—84.2\% of the inter-human baseline. Agreement was 
highest on objective dimensions such as verbosity (93.2\% of human-level 
agreement) and lower on subjective dimensions such as technical correctness 
(86.7\% of human-level agreement). Full methodology, annotation 
protocols, and judge–human comparison details are provided in 
Appendix~\ref{sec:expert-validation-appendix}.

\subsubsection{Error Category Analysis}

To better characterize how models fail across different problem types, we categorized the 310 cases from our human annotation study (Section~\ref{sec:expert-validation-appendix}) into seven coarse error types using LLM-assisted classification with Sonnet~3.7. Logic errors (28.8\%) and environment configuration issues (28.5\%) dominate the dataset, while performance-related issues exhibit the highest annotator–judge agreement (75.0\%). The full error taxonomy, per-model breakdowns, and implications for future benchmark design are provided in Appendix~\ref{sec:error-category-appendix}.

\subsubsection{Satisfaction Condition Validation}
We evaluated automatically extracted satisfaction conditions through annotations by 21 professional contractors (mean 5.1 years programming experience). Across 663 conditions from 70 randomly sampled issues, 86.3\% were judged accurate but only 65.7\% complete, indicating high precision with conservative recall. Our pipeline favors reliable extractions over exhaustive coverage. Detailed annotation guidelines and per-language statistics are in Appendices~\ref{sec:annotation} and~\ref{sec:human-alignment-appendix}.

\section{Limitations}
\label{sec:limitations}

While CodeAssistBench (CAB) offers a realistic and scalable framework for evaluating multi-turn programming assistance, several limitations remain, including conservative condition extraction, limited evaluation coverage, templated user behavior, and language scope constraints. Additionally, our LLM-based judge achieves 65.92\% agreement with human annotators (84.2\% of the 78.28\% inter-human baseline), suggesting multi-turn evaluation introduces alignment challenges compared to single-turn settings; see Appendix~\ref{sec:limitations-appendix} for full discussion.

\section{Conclusion and Future Work}

Our study introduces CodeAssistBench (CAB), a fully automated benchmark for evaluating multi-turn programming assistance grounded in real-world developer interactions. Unlike prior benchmarks that emphasize single-turn or synthetic tasks, CAB simulates realistic conversations over full codebases using containerized environments and issue-specific satisfaction criteria. Through extensive experiments, we find that state-of-the-art language models struggle with complex, multi-turn dialogues - particularly on recent repositories. CAB's automated pipeline enables scalable dataset expansion while supporting rigorous, reproducible evaluation across languages, project types, and time periods, offering a more faithful measure of model capabilities than existing benchmarks.

Several promising directions could extend CAB’s impact: improving satisfaction-condition extraction to balance precision and recall, expanding language coverage beyond the current seven languages, incorporating more sophisticated user-simulation strategies, and designing specialized metrics for assistance categories (Appendix~\ref{sec:api-testing-appendix}). We hope CAB enables deeper insights into current system limitations and guides the development of future AI assistants that more effectively support real-world software engineering workflows. We further discuss potential societal benefits and risks in Appendix~\ref{sec:impacts}.

\section*{Acknowledgments}
We thank Karthik N. R. (nrkrk@amazon.com) from AWS AI Labs for improving the maintainability and usability of our codebase, and the broader AWS AI team for valuable discussions and feedback that shaped this work.

\bibliographystyle{unsrtnat}
\bibliography{references}

\appendix
\section{LLM Filtering Prompts}
\label{appendix:filter-prompts}

\subsection{Issue-Level Relevance Filter}
\label{sec:issuefilter}
To identify issues suitable for benchmarking, we use a 7-question prompt to assess resolution status, specificity, clarity, and safety. Below is the exact prompt provided to the LLM:

\begin{quote}
\begin{lstlisting}[basicstyle=\ttfamily\small, frame=single, label={lst:issue-prompt}]
Please evaluate the following GitHub issue and its comments:

    Title: {title}

    Author: {author}

    Body:
    {body}

    Comments:
    {comments}

    Based on this conversation, please answer the following questions with Yes or No:
    1. Is the problem resolved by someone other than the author (not self-answered)?
    2. Does the conversation contain confirmation from the author that the problem has been resolved?
    3. Is the problem a specific technical issue (not a feature request, opinion, or open-ended question)?
    4. Is there a clear, definitive solution provided within the conversation?
    5. Can the solution be directly applied without requiring additional context or resources?
    6. Does the conversation contain any personally identifiable information (PII) such as Email addresses, Phone numbers, Physical addresses, Full names (beyond just GitHub usernames), Passwords or credentials, Personal identification numbers, IP addresses, or Any other sensitive personal information?
    7. Can this problem be reproduced and solved using the provided solution today (April 2025)?

Please provide your answers in the format:
    1. [Yes/No]
    2. [Yes/No]
    3. [Yes/No]
    4. [Yes/No]
    5. [Yes/No]
    6. [Yes/No]
    7. [Yes/No]
\end{lstlisting}
\end{quote}

The model responds with binary answers to each question, which we then parse to determine whether the issue is usable.

\subsection{Message-Level Comment Filter}
For message filtering, we construct the full conversation context and ask the LLM to identify non-contributory messages (e.g., "+1", "Thanks", "Bump"). A comment is retained unless explicitly marked for removal. The prompt follows:

\begin{quote}
\begin{lstlisting}[basicstyle=\ttfamily\small, frame=single, label={lst:comment-prompt}]
Analyze each comment in this GitHub issue conversation:

{conversation}

Your task is to identify ONLY comments that have ABSOLUTELY NO support-related value.
A comment should ONLY be removed if it falls into ALL of these criteria:
- Contains NO technical information
- Provides NO context about the issue
- Asks NO relevant questions (technical or process-related)
- Provides NO status updates or next steps
- Offers NO feedback on proposed solutions
- Contains NO clarifications about the user's situation or environment
- Has NO administrative or process value (like assigning work, requesting more info)

Examples of comments to remove:
1. Pure social messages at conversation end: "Thanks!", "Cool", "thumbs-up imoji"
2. Empty status updates: "+1", "Same issue", "Any updates?", "Bump" with no additional context
3. Completely off-topic discussions unrelated to the issue

IMPORTANT: Preserve comments that show the natural flow of support interaction. If a comment contains ANY support-related value, even if minimal or alongside thanks/acknowledgements, DO NOT remove it.

List ONLY the comment numbers that should be removed because they have absolutely no support-related value.
Format: 
NUMBERS: <comma-separated list of numbers>
EXPLANATION: <specific reasons why these comments add no support-related value>

If no comments should be removed, respond with:
NUMBERS: none
EXPLANATION: All comments contain some support-related value or context
\end{lstlisting}
\end{quote}

\subsection{Automated Dockerfile Synthesis and Repair}
\label{sec:dockerfile}

To ensure that every benchmarked issue can be built and tested in a fully self‑contained environment, we automatically \emph{(i)} generate an initial set of Dockerfile candidates and \emph{(ii)} iteratively refine any candidate that fails to build.

\paragraph{Phase 1: Candidate Generation.}
For each GitHub issue we collect a rich context bundle—repository URL, commit SHA, truncated issue body, \texttt{README}, a structure summary of the repo, up to two GitHub Actions workflows, and (if available) a reference Dockerfile from the same project.  
We then prompt the LLM to produce 5 self‑contained Dockerfiles, each of which must (1) install dependencies, (2) clone the repo at the specified commit, and (3) build the project without executing the user’s code.  
The system prompt fixes the LLM’s persona (“expert Docker engineer”), while the user prompt supplies the per‑issue context.  By requesting \emph{plain text only} (no Markdown fences) we can pipe the response directly to \texttt{docker build}.

\begin{lstlisting}[basicstyle=\ttfamily\small, frame=single, caption={Prompt for candidate generation}]
SYSTEM: You are an expert Docker engineer who creates Dockerfiles to build and validate GitHub issues.

Repository URL: {repo_url}
Title: {issue_title}
Commit SHA: {commit_sha}

Issue description (truncated to 3 kB):
{issue_body}

Context:
- README
- Repo structure summary
- GitHub workflow files (optional)  
- Reference/Original Dockerfile (optional)

Create a Dockerfile that
1. installs all dependencies in the issue,
2. clones the repository and checks out {commit_sha} or user given project version,
3. builds the project (no test or run commands)

IMPORTANT: Return only the raw Dockerfile content - no Markdown, no commentary.
\end{lstlisting}

\paragraph{Phase 2: Fault‑Directed Repair.}
If every candidate in Phase 1 fails to build, we capture the build log and feed it—together with the failing Dockerfile and the same context bundle—into a repair prompt.  
The LLM is asked to produce an improved Dockerfile that specifically addresses the observed errors.  We repeat this loop for up to three attempts or until a build succeeds.

\begin{lstlisting}[basicstyle=\ttfamily\small, frame=single, caption={Prompt for candidate repair}]
SYSTEM: You are an expert Docker engineer who specializes in fixing Dockerfiles that failed to build.

USER:
Repository URL: {repo_url}
Issue #: {issue_number}   Title: {issue_title}
Commit SHA: {commit_sha}

Failing Dockerfile:
{candidate_dockerfile}

Build error (truncated to 3 kB):
{build_error}

Provide a corrected Dockerfile that
1. removes the above error(s),
2. keeps the minimal environment needed to build,
3. follows the same constraints as the generation prompt.

IMPORTANT: Return only the raw Dockerfile content - no Markdown, no commentary.
\end{lstlisting}

\subsection{Satisfaction‑Condition Extraction}
\label{sec:satcond}

To evaluate whether an assistant’s reply \emph{actually} meets a developer’s needs, we first distill each GitHub issue thread into a small set of \textbf{user‑satisfaction conditions}—explicit criteria that any acceptable answer must fulfill. The extraction procedure is entirely automated and comprises two stages.

\paragraph{Stage 1: LLM‑based Extraction.}
We prompt a language model to extract these conditions from each conversation using two coordinated prompts: a \emph{system prompt} that defines the task, abstraction level, and response format, and a \emph{user prompt} that injects issue-specific content. The model returns a JSON object describing each condition along with a brief explanation.

\begin{quote}
\begin{lstlisting}[basicstyle=\ttfamily\small, frame=single, caption={System prompt for satisfaction-condition extraction}, label={lst:sat-system}]
You are an expert at analyzing GitHub issues and extracting user satisfaction conditions-the criteria by which any answer will be judged.

A satisfaction condition states WHAT the user needs, not HOW to implement it.

# Abstraction guide
TOO SPECIFIC - "Use numpy.where(...)"
GOOD LEVEL   - "Vectorized conditional selection"
TOO GENERIC  - "A working solution"

# Must-have properties
1. TRANSFERABLE  (solution-agnostic)
2. VERIFIABLE    (pass/fail is clear)
3. EVIDENCED     (grounded in user utterances)
4. NEED-FOCUSED  (problem, not implementation)

Return exactly this JSON:
{
  "satisfaction_conditions": [
    {
      "condition": "...",
      "explanation": "..."
    }
  ]
}
Do not wrap the JSON in markdown fences.
\end{lstlisting}
\end{quote}

\vspace{0.8em}

\begin{lstlisting}[basicstyle=\ttfamily\small, frame=single, caption={User prompt for satisfaction-condition extraction}, label={lst:sat-user}]
Given this GitHub conversation:

Title   : {title}
Author  : {author}
Question: {body}

Comments (chronological, at most 100):
{formatted_comments_json}

Extract every user satisfaction condition.
Remember: they are *criteria*, not solutions.

Output exactly one JSON object as described in the system prompt-no extra text.
\end{lstlisting}

\paragraph{Stage 2: Post‑hoc Verification.}
The raw JSON returned by the LLM is parsed and each condition is passed through a lightweight verifier that checks:
\begin{enumerate}
  \item the text parses as valid JSON;
  \item each entry includes both a \texttt{condition} and an \texttt{explanation};
  \item the explanation quotes or paraphrases evidence found in the thread.
\end{enumerate}

Conditions that fail verification are discarded. Issues with no surviving valid conditions are excluded from the benchmark.

\subsection{LLM judge}
\label{sec:judge-maintainer}

To assess the quality of the maintainer's answer in each conversation, we employ a structured LLM prompt that simulates a judge agent. The agent is provided with (i) the original GitHub issue (title, body, and comments), (ii) the set of user satisfaction conditions, (iii) the maintainer’s answer to be evaluated, and (iv) optional Docker validation results. The LLM is instructed to evaluate the answer along three axes: technical correctness, alignment with user satisfaction conditions, and verbosity.

\paragraph{Evaluation Prompt.} The judge agent uses the prompt shown in Listing~\ref{lst:judge-prompt}:
\begin{quote}
\begin{lstlisting}[basicstyle=\ttfamily\small, frame=single, caption={Prompt for LLM judge}, label={lst:judge-prompt}]
You are a judge evaluating the maintainer's answer to a user's technical question.

Your task is to determine if the maintainer's answer is:
1. TECHNICALLY CORRECT
2. SATISFIES USER CONDITIONS
3. APPROPRIATE VERBOSITY

IMPORTANT: For Docker-related issues:
- The answer must be technically correct AND
- The Docker build/test process must succeed

If Docker validation fails (Success: False), the answer is considered INCORRECT.

Provide your evaluation in this format:

TECHNICAL CORRECTNESS: [CORRECT / PARTIALLY CORRECT / INCORRECT]

ALIGNMENT SCORE: X/Y CONDITIONS MET (Z%)

CONDITION 1: [TRUE / FALSE] <brief description>
...

VERBOSITY ASSESSMENT: [TERSE / APPROPRIATE / VERBOSE]

VERDICT: [CORRECT / PARTIALLY CORRECT / INCORRECT]

KEY ISSUES:
- Issue 1
- Issue 2

REASONING:
Detailed explanation of correctness and alignment.
\end{lstlisting}
\end{quote}

\paragraph{Inputs.} The judge agent receives:
\begin{itemize}
    \item The full conversation (title, question body, comments),
    \item User satisfaction conditions,
    \item Maintainer’s generated answer,
    \item Docker validation logs (if applicable).
\end{itemize}

\paragraph{Verdict Criteria.} The final judgment is based on:
\begin{itemize}
    \item \textbf{Correct:} Fully accurate and satisfies \textit{all} user conditions.
    \item \textbf{Partially Correct:} Minor technical flaws or partial condition satisfaction.
    \item \textbf{Incorrect:} Major errors, unmet conditions, or failed Docker validation.
\end{itemize}

\paragraph{Post-Processing.} The LLM's output is parsed to extract:
\begin{itemize}
    \item Technical correctness,
    \item Number of conditions satisfied,
    \item Verbosity assessment,
    \item Final verdict,
    \item Key issues and reasoning.
\end{itemize}

\section{Expert Validation: Detailed Methodology and Results}
\label{sec:expert-validation-appendix}

\subsection{Validation Methodology}

\paragraph{Dataset Construction.}
We initially sampled 200 unique GitHub issues from real-world repositories across seven programming languages for expert validation. To ensure annotation quality, we filtered out 24 issues containing non-English content, as all annotators were English speakers, resulting in a final set of 176 unique issues. 

We then generated agent responses using two state-of-the-art models for these 176 issues: ChatGPT 4.1 Mini and Claude Sonnet 3.7. Initially, both models generated responses for most issues (134 issues received responses from both models to enable comparative analysis). However, not all AI-generated responses were suitable for human annotation. We excluded responses where the agent output contained unicode character encoding issues or null fields in critical parts, which would have prevented annotators from accurately evaluating the responses. Specifically, 40 GPT responses and 38 Sonnet responses were excluded due to these data quality issues.

After this filtering, we retained 310 usable agent responses for human evaluation: 154 from ChatGPT 4.1 Mini and 156 from Claude Sonnet 3.7. Each of these 310 responses was independently evaluated by two expert annotators, yielding 620 total annotation records. The final language-wise distribution of validated agent responses per model is shown in Table~\ref{tab:validation-distribution}.

\begin{table}[h]
\centering
\caption{Language distribution of validated agent responses}
\label{tab:validation-distribution}
\begin{tabular}{lcc}
\hline
\textbf{Language} & \textbf{ChatGPT 4.1 Mini} & \textbf{Claude Sonnet 3.7} \\
\hline
Python      & 41 & 42 \\
JavaScript  & 30 & 31 \\
C++         & 25 & 25 \\
C\#         & 22 & 21 \\
TypeScript  & 17 & 18 \\
C           & 13 & 13 \\
Java        &  6 &  6 \\
\hline
\textbf{Total} & \textbf{154} & \textbf{156} \\
\hline
\end{tabular}
\end{table}

\paragraph{Annotation Protocol.}
Two senior software engineers (Human1 and Human2) with 3+ years of experience each independently evaluated all 310 agent responses. Critically, annotators used the \textit{exact same evaluation prompts and criteria} as the LLM judges to ensure fair comparison and minimize methodological confounds. Each response was evaluated across three dimensions:

\begin{itemize}
    \item \textbf{Technical Correctness}: Does the response accurately address the user's technical question?
    \item \textbf{Verbosity}: Is the response appropriately concise, verbose, or terse?
    \item \textbf{Overall Verdict}: Is the response correct, partially correct, or incorrect?
\end{itemize}

Annotators worked independently without communication, and each agent response was evaluated by both annotators, yielding 620 annotation records (310 responses $\times$ 2 annotators).

\subsection{Inter-Rater Reliability}

The two expert annotators demonstrated substantial overall agreement at 78.28\% (728/930 judgments across all dimensions; Cohen's $\kappa = 0.68$), establishing a robust baseline for evaluating LLM judge performance. According to established guidelines~\cite{landis1977measurement}, $\kappa$ values between 0.61--0.80 indicate substantial agreement, confirming that our annotation protocol achieves a high level of consistency. Agreement varied across evaluation dimensions, as shown in Table~\ref{tab:inter-rater}.

\begin{table}[h]
\centering
\caption{Inter-rater reliability between Human1 and Human2}
\label{tab:inter-rater}
\begin{tabular}{lccc}
\hline
\textbf{Dimension} & \textbf{Agreement} & \textbf{Percentage} & \textbf{Cohen's $\kappa$} \\
\hline
Verbosity                 & 263/310 & 84.84\% & 0.77 \\
Technical Correctness     & 233/310 & 75.16\% & 0.63 \\
Verdict                   & 232/310 & 74.84\% & 0.62 \\
\hline
\textbf{Overall}          & \textbf{728/930} & \textbf{78.28\%} & \textbf{0.68} \\
\hline
\end{tabular}
\end{table}

This variation reveals that while response style (verbosity) is relatively objective and unambiguous ($\kappa = 0.77$, substantial agreement), assessing technical accuracy requires nuanced domain expertise and can yield different but equally defensible judgments ($\kappa = 0.62\text{--}0.63$, moderate to substantial agreement). According to established guidelines~\cite{landis1977measurement}, $\kappa$ values between 0.61--0.80 indicate substantial agreement, and our results fall within this range across all dimensions. This level of agreement demonstrates that even experienced practitioners may weigh correctness, completeness, and stylistic factors differently when evaluating subjective technical tasks.

Our 78.28\% agreement rate and Cohen's $\kappa$ of 0.68 represent substantial inter-rater reliability, validating both our annotation protocol and confirming the inherent subjectivity in evaluating code quality and assistance effectiveness.

\subsection{LLM-Expert Agreement}

We compared automated LLM judge evaluations against both expert annotators to assess the reliability of LLM-as-judge for code assistance evaluation. Table~\ref{tab:llm-expert} summarizes the agreement rates.

\begin{table}[h]
\centering
\caption{Agreement rates between LLM judges and expert annotators}
\label{tab:llm-expert}
\begin{tabular}{lcccc}
\hline
\textbf{Judge Pair} & \textbf{Overall} & \textbf{Technical} & \textbf{Verbosity} & \textbf{Verdict} \\
\hline
Human1 vs Human2    & 78.28\% & 75.16\% & 84.84\% & 74.84\% \\
LLM vs Human1       & 70.22\% & 65.16\% & 79.03\% & 66.45\% \\
LLM vs Human2       & 61.61\% & 53.87\% & 74.84\% & 56.13\% \\
\hline
\end{tabular}
\end{table}

The LLM judges achieved 70.22\% agreement with Human1, representing 89.7\% of the inter-human baseline (78.28\%), and 61.61\% with Human2 (78.7\% of baseline). The average LLM-human agreement of 65.92\% reaches 84.2\% of the inter-human agreement level, demonstrating that LLM judges approach human-level evaluation capabilities while maintaining scalability advantages. These results reveal several important findings:

\paragraph{Performance on Objective vs. Subjective Metrics.}
LLM judges achieve their highest agreement with human experts on verbosity assessment (74--79\%), where evaluation criteria are most clearly defined and objective. Agreement decreases for technical correctness (54--65\%), where deep domain expertise and contextual understanding become critical. The 79.03\% LLM-Human1 agreement on verbosity (93.2\% of the 84.84\% human baseline) demonstrates that automated evaluation excels at surface-level characteristics but requires improvement for nuanced technical assessment.

\paragraph{Variability in Human Standards.}
The substantial 8.61 percentage point difference in LLM agreement between Human1 (70.22\%) and Human2 (61.61\%) indicates that individual annotator strictness, expertise emphasis, or judgment criteria significantly impact evaluation outcomes. This variability—comparable to the 8.06 point gap between Human1-LLM (70.22\%) and Human1-Human2 (78.28\%)—highlights the importance of multi-annotator validation and suggests that single-annotator studies may yield inconsistent conclusions. Our dual-annotator approach provides more reliable ground truth than typical single-expert evaluation.

\section{Extended Discussion of Limitations}
\label{sec:limitations-appendix}

While CodeAssistBench (CAB) provides a realistic and scalable framework for evaluating multi-turn programming assistance, it has several limitations.

First, our satisfaction condition extraction prioritizes precision over recall. Although 86.3\% of extracted conditions were judged accurate by human annotators, only 65.7\% were complete, suggesting that models may be unfairly penalized for omitting criteria not fully captured by our automated pipeline.

Second, evaluation is performed on a sampled subset of 544 issues (from a pool of 3,286), constrained by computational cost. This may skew results away from rare or edge-case conversations.

Third, the simulated user uses BM25-matched historical responses to simulate follow-ups. While grounded in real-world interactions, this approach may underrepresent the full diversity of developer behaviors, especially in ambiguous or exploratory contexts.

Fourth, CAB only includes issues with successfully synthesized Docker environments. This excludes legacy or atypically configured projects, introducing a bias toward actively maintained, modern repositories with standard build systems.

Fifth, our evaluation targets only seven programming languages and repositories with permissive open-source licenses. The benchmark does not currently assess proprietary codebases, enterprise workflows, or other language ecosystems (e.g., Rust, Go, Swift).

Sixth, we fixed generation temperatures to promote reproducibility: temperature=0 for conversation responses and temperature=0.7 for Dockerfile generation (to encourage diversity across five sampled candidates). While this setup ensures consistency and comparability, it may not reflect each model’s optimal decoding parameters for task performance.

Seventh, we do not report error bars or statistical significance tests. While aggregate model trends are clear, fine-grained comparisons—especially between close-performing models—should be interpreted cautiously.

Lastly, the 12.36 percentage point gap between our LLM judge's agreement with human annotators (65.92\%) and the inter-human baseline (78.28\%) represents a fundamental limitation of automated multi-turn evaluation. This gap is particularly pronounced for subjective dimensions like technical correctness (54--65\% LLM-human agreement) compared to objective dimensions like verbosity (74--79\%). 

We hypothesize that multi-turn conversations introduce several sources of variance that make alignment harder than single-turn evaluation:
\begin{enumerate}
    \item \textbf{Accumulated context interpretation}: As conversations progress, the accumulated context can be interpreted differently by human and LLM judges. A response that seems adequate given early turns may be judged differently when later clarifications reveal the user's true intent.
    \item \textbf{Credit attribution ambiguity}: In multi-turn dialogues, it becomes unclear which turn's contribution should be credited for ultimate success or failure. Human annotators may attribute success to the final clarifying response, while LLM judges may weight earlier turns more heavily.
    \item \textbf{Trajectory-dependent evaluation}: The combinatorial space of possible conversation trajectories makes it difficult to establish consistent evaluation criteria. Two conversations reaching the same endpoint via different paths may receive different judgments depending on the intermediate steps.
    \item \textbf{Error propagation}: Early misunderstandings or incomplete responses can cascade through subsequent turns, making it challenging to isolate which specific failure led to an incorrect final assessment.
\end{enumerate}

This finding has important implications for the broader research community working on LLM-as-judge methods and multi-turn dialogue evaluation. We suggest several directions for future work to address this limitation:
\begin{itemize}
    \item \textbf{Turn-level annotations}: Rather than evaluating entire conversations holistically, decomposing evaluation into per-turn assessments may reduce variance and improve alignment.
    \item \textbf{Trajectory-based metrics}: Developing metrics that explicitly account for conversation dynamics and progression, rather than treating dialogues as atomic units.
    \item \textbf{Consensus-based judging}: Using multiple LLM judges with ensemble methods to reduce individual model biases, similar to how we used multiple human annotators.
    \item \textbf{Calibrated confidence scores}: Training judges to output calibrated confidence scores that reflect uncertainty in multi-turn contexts, enabling downstream systems to weight judgments appropriately.
\end{itemize}

We leave the systematic investigation of these approaches to future work, but note that improving multi-turn evaluation reliability remains critical for advancing conversational code assistance research.

We believe addressing these limitations—by improving condition recall, scaling human annotation, diversifying simulation strategies, and exploring a broader language and tooling spectrum—will further enhance CAB’s utility as a long-term benchmark for conversational coding agents.

\section{Broader Impacts}
\label{sec:impacts}
CodeAssistBench (CAB) has the potential to advance the development of safer and more capable AI programming assistants, particularly in realistic, multi-turn developer scenarios. By surfacing failure modes and limitations in current models, CAB encourages the creation of tools that better align with developer needs, which may enhance productivity and learning. However, there are also potential negative impacts. For instance, if developers rely too heavily on inaccurate agent suggestions in real-world settings, this could introduce subtle bugs or security issues. Additionally, the use of scraped GitHub data—even when filtered—may raise concerns around privacy or attribution if not carefully managed. We encourage responsible use of CAB, including proper consent for dataset extension and monitoring downstream usage of models trained or benchmarked using this framework.

\section{Hardware Requirements and Runtime Environment}
\label{sec:requirements}

Most stages of the CodeAssistBench (CAB) pipeline—including dataset generation, model evaluation, and human alignment analysis—are driven by API-based interactions with GitHub and commercial LLM endpoints. As a result, these stages require minimal local compute resources and can be executed efficiently on standard cloud instances or modest local machines.

The only hardware-intensive component is the Dockerfile synthesis and validation phase. This stage involves generating candidate Dockerfiles and building them in parallel across hundreds of repositories to ensure correctness and reproducibility. Due to the storage and memory demands of large-scale container builds, we recommend using machines equipped with at least 1TB of local storage and 32GB of memory.

All experiments were conducted on AWS \texttt{g5.16xlarge} instances, each with 1 NVIDIA A10G GPU, 64 vCPUs, and 256GB of RAM. While GPU acceleration was not required (as all model queries were served via commercial APIs), the instance's large memory and CPU capacity were beneficial during Docker-based validation. The total estimated compute usage was approximately 100 CPU-hours, with dataset generation taking around 3 days and experimental runs completing in approximately 4 days.

\section{Docker Build Details}
\label{sec:docker-build-details}

Most build failures in All-Time repositories stemmed from outdated dependencies or unavailable components, not from limitations of the LLM pipeline itself. To address this, CAB inferred project-specific requirements—such as relevant library versions, operating systems, and toolchains—by analyzing both the issue's content and its creation timestamp, then tailored environment reconstruction accordingly via custom Dockerfile generation. The higher success rate in Recent repositories (97.6\% vs. 78.2\%) reflects better availability of current dependencies and more standardized build configurations in modern projects.

\section{Supplementary Material}
\label{sec:supplementary}

\begin{figure}[ht]
\centering
\includegraphics[width=0.95\linewidth]{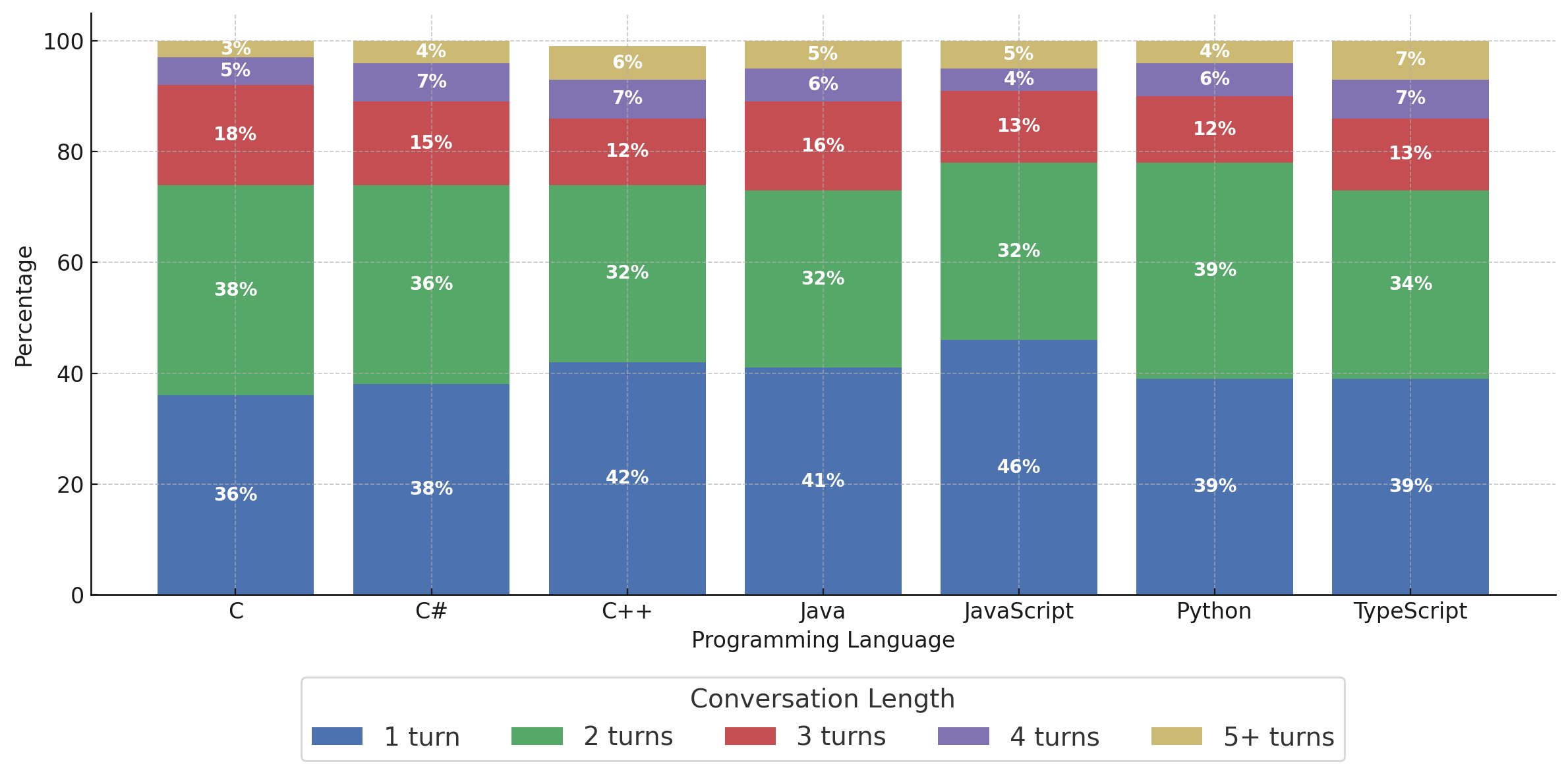}
\caption{Distribution of GitHub issue conversation lengths by programming language. Each turn corresponds to a maintainer response; for example, a 1-turn conversation consists of a user question and a single maintainer reply, while longer conversations reflect additional back-and-forth exchanges.}
\label{fig:conv-length-distribution}
\end{figure}

\begin{table}[ht]
\centering
\caption{Docker environment build success rates by programming language}
\label{tab:docker_builds}
\begin{tabular}{lrrr}
\toprule
\textbf{Language} & \textbf{Environment-dependent} & \textbf{Successful Builds} & \textbf{Success Rate} \\
\midrule
C & 17 & 17 & 100\% \\
C\# & 30 & 21 & 70\% \\
C++ & 53 & 28 & 53\% \\
Java & 58 & 52 & 90\% \\
JavaScript & 34 & 33 & 97\% \\
Python & 90 & 77 & 86\% \\
TypeScript & 12 & 10 & 83\% \\
\midrule
\textbf{Total} & 294 & 238 & 81\% \\
\bottomrule
\end{tabular}
\end{table}

\begin{figure}[ht]
  \centering
  \includegraphics[width=\linewidth]{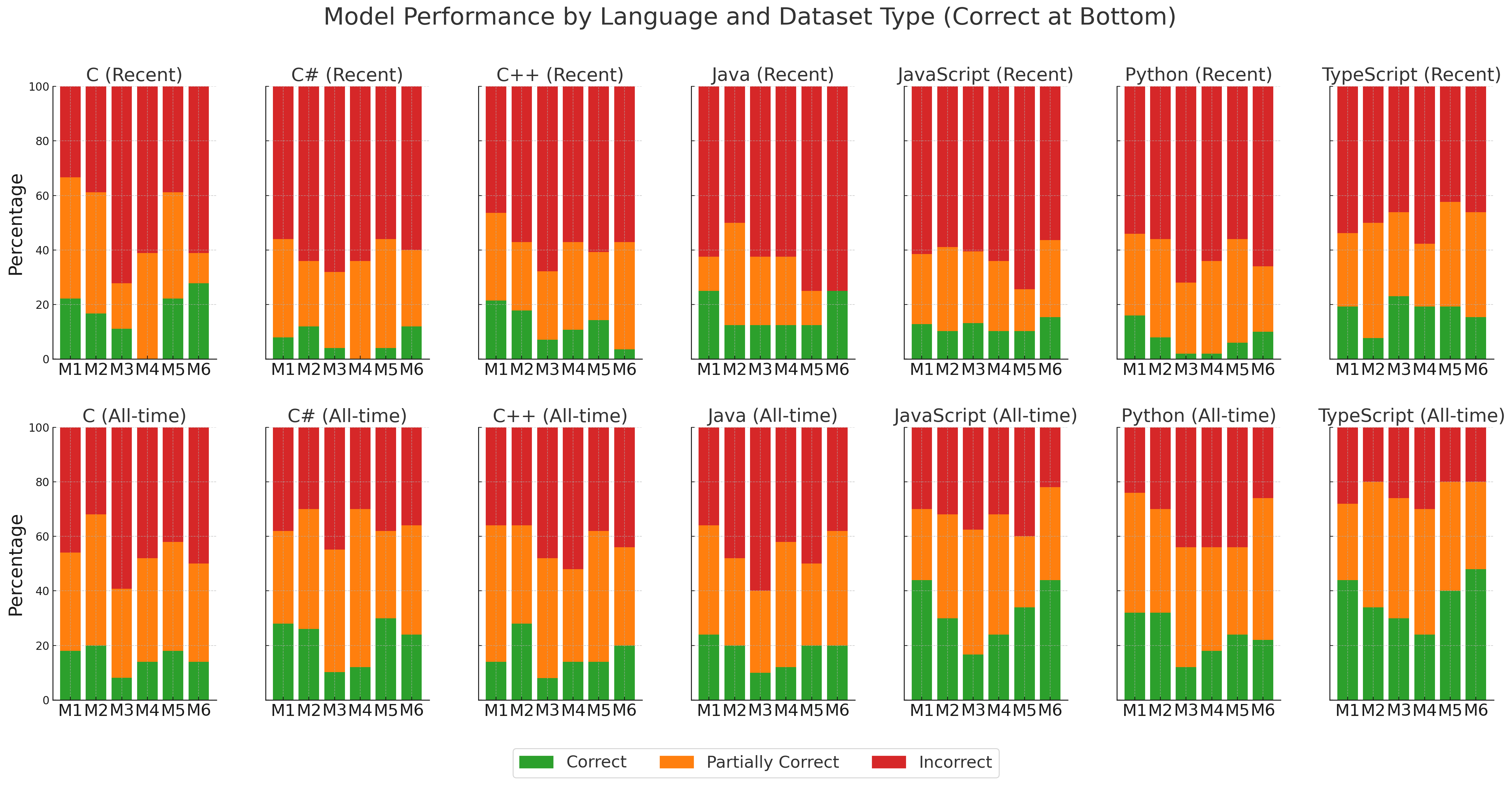}
  \caption{
  Cumulative comparison of model performance across seven programming languages. 
  Each bar represents a model's prediction outcome distribution on GitHub issues, broken down into 
  \textbf{Correct (green)}, \textbf{Partially Correct (orange)}, and \textbf{Incorrect (red)} responses. 
  The top row corresponds to evaluations on \textbf{recent repositories}, while the bottom row shows results on \textbf{all-time repositories}.
  Models are labeled as M1–M6 in the following order: 
  \textbf{M1}: ChatGPT 4.1 Mini, 
  \textbf{M2}: DeepSeek R1, 
  \textbf{M3}: Llama 3.3 70B, 
  \textbf{M4}: Haiku 3.5, 
  \textbf{M5}: Sonnet 3.7, 
  \textbf{M6}: Sonnet 3.7 (Think Mode).
  }
  \label{fig:languages}
\end{figure}

\begin{figure}[ht]
  \centering
  \includegraphics[width=\linewidth]{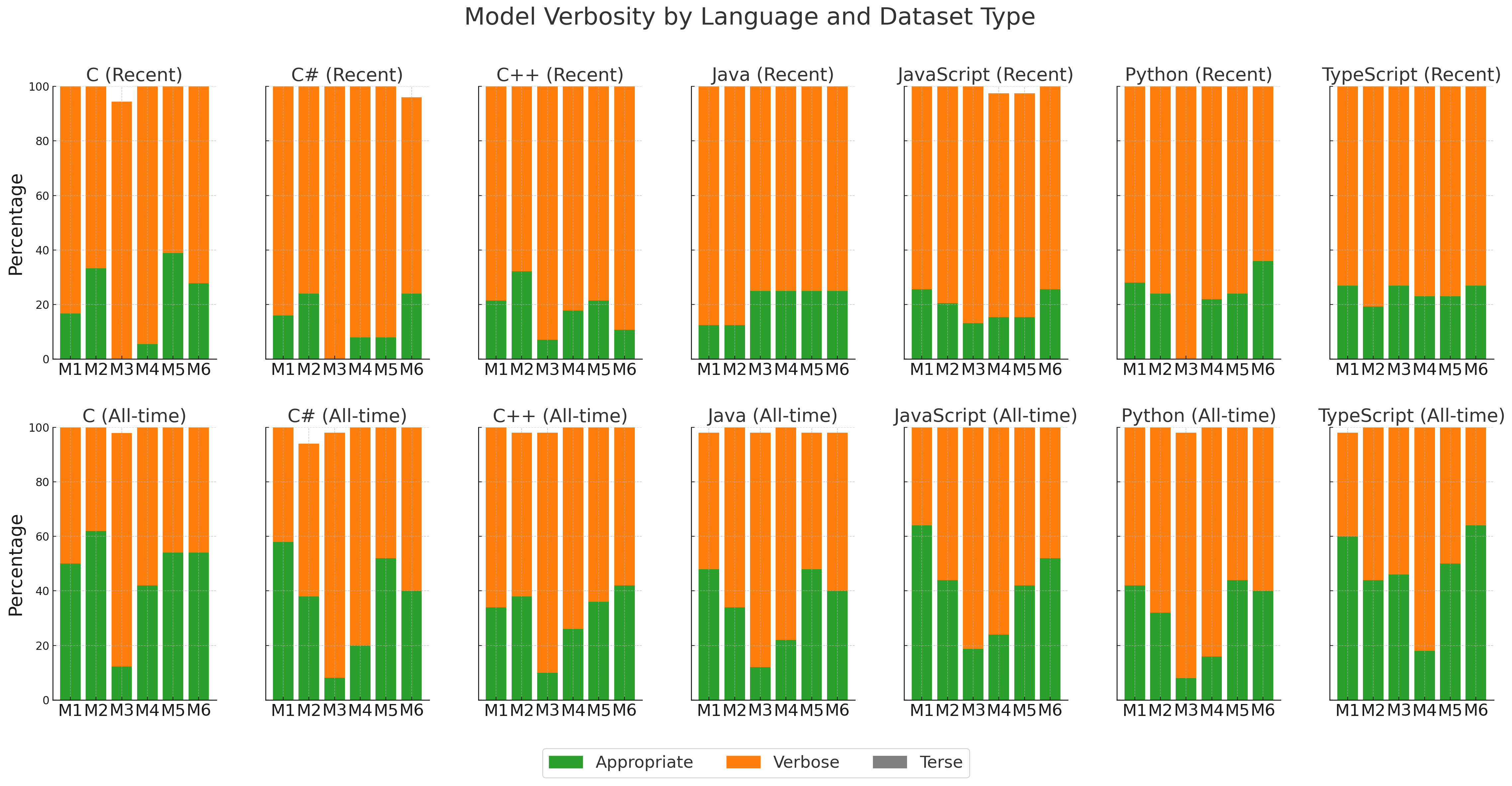}
  \caption{
  Cumulative verbosity distribution of model responses across seven programming languages.
  Each bar shows the proportion of responses classified as \textbf{Appropriate} (green), \textbf{Verbose} (orange), and \textbf{Terse} (gray, rarely observed).
  The top row corresponds to evaluations on \textbf{recent repositories}, while the bottom row shows results on \textbf{all-time repositories}.
  Models are labeled as M1–M6 in the following order: 
  \textbf{M1}: ChatGPT 4.1 Mini, 
  \textbf{M2}: DeepSeek R1, 
  \textbf{M3}: Llama 3.3 70B, 
  \textbf{M4}: Haiku 3.5, 
  \textbf{M5}: Sonnet 3.7, 
  \textbf{M6}: Sonnet 3.7 (Think Mode).
  Most models tend to generate verbose responses, with only a few models (notably M6) achieving higher rates of appropriately concise answers.
  }
  \label{fig:verbosity}
\end{figure}

\newpage
\section{Satisfaction Condition Validation Details}
\label{sec:human-alignment-appendix}

We conducted a targeted evaluation to assess the accuracy and completeness of automatically extracted user satisfaction conditions. We randomly sampled 10 GitHub issues per language across seven programming languages and collected annotations from three independent human raters. Each annotator evaluated two aspects: \textbf{Condition Accuracy} and \textbf{Condition Completeness}.  Across 210 records, annotators reviewed a total of 663 satisfaction conditions. The detailed per‐language breakdown is shown in Table~\ref{tab:alignment-metrics}.

\begin{table}[ht!]
\centering
\caption{Human Alignment Evaluation: Accuracy, Recall, and Additional Condition Needs per Language}
\label{tab:alignment-metrics}
\resizebox{\textwidth}{!}{%
\begin{tabular}{lcccccc}
\toprule
\textbf{Language}  & \textbf{Total Conditions} & \textbf{Correct} & \textbf{Incorrect} & \textbf{Accuracy} & \textbf{Recall} & \textbf{Additional Needed (\%)} \\
\midrule
C           &  87 &  80 &   7 & 91.95\% & 96.67\% & 3.3\%  \\
C++         &  96 &  94 &   2 & 97.92\% & 26.67\% & 73.3\% \\
C\#         &  96 &  68 &  28 & 70.83\% & 70.00\% & 30.0\% \\
Java        & 102 &  87 &  15 & 85.29\% & 86.67\% & 13.3\% \\
JavaScript  &  93 &  80 &  13 & 86.02\% & 53.33\% & 46.7\% \\
Python      & 102 &  79 &  23 & 77.45\% & 66.67\% & 33.3\% \\
TypeScript  &  87 &  84 &   3 & 96.55\% & 60.00\% & 40.0\% \\
\midrule
\textbf{Overall} & 663 & 572 &  91 & 86.27\% & 65.71\% & 34.3\% \\
\bottomrule
\end{tabular}%
}
\end{table}

\paragraph{Annotator Agreement.}
We computed pairwise agreement and Cohen’s Kappa across the three annotators. The average percentage overlap was 82.39\%, with some pairs reaching perfect agreement (100.00\%, $\kappa=1.0$) and others showing low or negative $\kappa$, highlighting variability in interpretation and the need for further calibration in future rounds.

\section{Human Annotation Instructions}
\label{sec:annotation}

This section provides the full instructions shown to human annotators for evaluating the quality of user satisfaction conditions in CodeAssistBench (CAB). These instructions were designed to ensure consistency, transparency, and alignment with user intent during the evaluation process.

\subsection*{Ethics and Compensation}
Annotators were external contractors who provided informed consent prior to participation. They were compensated at or above the local minimum wage. The task involved minimal risk, and no personally identifiable or sensitive information was collected.

\subsection*{Task Overview}
This annotation task involves evaluating user satisfaction conditions automatically extracted from GitHub issue threads. A satisfaction condition describes **what** the user needs from the maintainer’s response—not **how** to implement it. These conditions should reflect the user’s goals, constraints, or expectations based on the full conversation.

Each annotator was asked to review 70 GitHub issues, with three annotators assigned per issue. A calibration phase with 10 examples was used to ensure alignment before the full task began.

\subsection*{Annotation Interface}
Annotators were presented with the following information for each task:

\begin{itemize}
  \item Issue title and initial question
  \item Full GitHub thread, including all comments and author roles
  \item The list of model-generated satisfaction conditions
\end{itemize}

\subsection*{Annotator Instructions (Verbatim)}
The following is the exact text shown to annotators:

\begin{quote}
\textbf{Overview}

This project focuses on verifying user satisfaction conditions automatically extracted from GitHub issue conversations using a language model. A satisfaction condition describes what the user needs from the maintainer’s response—not how it is implemented.

You will review GitHub issues and evaluate each proposed satisfaction condition for:
- \textbf{Correctness}: Does the condition accurately reflect user needs?
- \textbf{Completeness}: Are there missing conditions that should be included?

\textbf{For each issue, you will be given:}
- The original GitHub title and user question
- All follow-up comments (including roles and timestamps)
- The model-generated satisfaction conditions

\textbf{Your tasks:}
1. For each listed condition:
    - Mark it as \texttt{Correct} or \texttt{Incorrect}
    - Provide a short justification (e.g., “User confirmed this need in a follow-up comment.”)
2. Suggest any \textbf{missing conditions} that should be added, with justification.

\textbf{Annotation Format:}

\begin{verbatim}
{
  "original_conditions": [
    {
      "condition": "string",
      "judgment": "Correct" | "Incorrect",
      "justification": "string"
    }
  ],
  "additional_conditions (if any)": [
    {
      "condition": "string",
      "justification": "string"
    }
  ]
}
\end{verbatim}

\textbf{Levels of Abstraction for Conditions:}

Avoid overly specific or overly generic phrasing. Good conditions should reflect the \textit{what}, not the \textit{how}.

Too Specific (Avoid):
- "Use numpy.where(condition, x, y)"
- "Set max\_depth=5 in the RandomForest constructor"

Good Abstraction Level:
- "A vectorized approach to conditional element selection"
- "Guidance on suitable tree depth settings"

Too Generic (Avoid):
- "A working solution"
- "Help with the problem"

\textbf{Criteria for a Good Condition:}
- \textbf{Transferable}: Not tied to a specific implementation
- \textbf{Verifiable}: Can be clearly judged as satisfied or not
- \textbf{Evidenced}: Grounded in what the user said or implied
- \textbf{Needs-Focused}: Describes the user’s goal/problem—not the method
\end{quote}

\subsection*{Example Annotation}

Below is an anonymized real example used during calibration:

\textbf{Issue Title:} “How can I play RTSP stream without audio codecs?”

\textbf{User Post:}
\textit{How can I play RTSP stream without audio codecs? I need only video. I can't start watching because the camera uses G.711 for audio.}

\textbf{Model-Generated Conditions:}
\begin{itemize}
  \item Explanation of how the player handles unsupported audio codecs
  \item Confirmation that video playback is possible without audio codec support
\end{itemize}

\textbf{Annotator Output:}

\begin{lstlisting}
{
  "original_conditions": [
    {
      "condition": "Explanation of how the player handles unsupported audio codecs",
      "judgment": "Correct",
      "justification": "User asked about G.711 and the maintainer explained it would be dropped automatically."
    },
    {
      "condition": "Confirmation that video playback is possible without audio codec support",
      "judgment": "Correct",
      "justification": "The maintainer confirmed that video-only playback is supported."
    }
  ],
  "additional_conditions": []
}
\end{lstlisting}

\subsection*{Quality Control}

Each annotator completed a 10-example calibration round. Only those with consistent agreement advanced to the full task. Inter-annotator agreement statistics are reported in Appendix~\ref{sec:human-alignment-appendix}.

\subsection*{Annotator Backgrounds}

A total of 21 annotators participated in the human evaluation task. All annotators had at least 2 years of programming experience (mean: 5.1 years), with backgrounds spanning industry roles in software engineering, quality assurance, and backend development.

\begin{itemize}
  \item \textbf{Languages covered:} JavaScript/TypeScript, Python, Java, C\#
  \item \textbf{Common frameworks:} React, Angular, Spring Boot, Django, .NET Core
  \item \textbf{Cloud \& infra skills:} AWS, SQL, HTML/CSS
  \item \textbf{Review process:} Each GitHub issue was annotated by 3 independent reviewers.
\end{itemize}

A full table of individual annotator experience is available upon request but omitted from the appendix to preserve readability.

\section{Synthetic Dataset Ablation Study}
\label{sec:synthetic-ablation-appendix}

To investigate whether the observed performance gap between recent and all-time repositories stems from AI-generated code characteristics versus training data knowledge limitations, we conducted an ablation study using a synthetic dataset.

\subsection{Methodology}

We randomly selected 50 problems from the recent repository subset and generated fully synthetic reproductions using a maintainer agent backed by Claude Sonnet 3.7. For each issue, the agent was run iteratively until it confirmed the issue was fully reproducible within the generated codebase. Each synthetic repository maintained similar complexity (20-28 files, 1,500+ lines of code) while preserving the identical bugs and satisfaction conditions from the original issues. The synthetic repositories were designed to use programming language versions and library versions that fall within the model's training knowledge cutoff.

We initially hypothesized that AI-generated code characteristics might be a primary contributor to the observed low accuracy on recent repositories. However, during repository generation, the maintainer agent naturally produced extensive documentation as part of its workflow to ensure full reproducibility, resulting in synthetic repositories with significantly better documentation coverage than typical real-world repositories.

\subsection{Results}

The synthetic dataset evaluation revealed a striking finding: Claude 3.7 Sonnet achieved \textbf{74\% accuracy} on the synthetic repositories, substantially higher than the 11.34\% accuracy observed on the recent repository subset (Table~\ref{tab:combined-accuracy-corrected}).
 
\begin{table}[h]
\centering
\caption{Performance comparison across repository types (Sonnet 3.7)}
\label{tab:synthetic-comparison-appendix}
\begin{tabular}{lcccc}
\toprule
\textbf{Dataset Type} & \textbf{Correct} & \textbf{Partial} & \textbf{Incorrect} & \textbf{Total Accuracy} \\
\midrule
All-time repos & 25.71\% & 35.43\% & 38.86\% & 61.14\% \\
Recent repos & 11.34\% & 30.93\% & 57.73\% & 42.27\% \\
Synthetic repos & \textbf{74.0\%} & 16.0\% & 10.0\% & \textbf{90.0\%} \\
\bottomrule
\end{tabular}
\end{table}

\subsection{Analysis and Implications}

The superior performance on synthetic code compared to recent repositories provides compelling evidence that \textbf{the performance gap is primarily driven by training data limitations rather than inherent properties of AI-generated code}. Several key factors explain this finding:

\paragraph{Library Version Knowledge} The most significant factor contributing to failures in recent repositories was the use of library versions and programming language features released after the model's training data cutoff. Recent repositories frequently employ:
\begin{itemize}
    \item Latest library APIs introduced after the training cutoff
    \item Modern language features from recent standard updates
    \item Newly released frameworks and their evolving best practices
\end{itemize}

In contrast, the synthetic repositories were constructed using language and library versions that fall well within the model's training knowledge, enabling the model to leverage its full understanding of these technologies.

\paragraph{Repository Complexity} Synthetic repositories, while maintaining realistic complexity (1,500+ LOC, professional structure), were necessarily smaller and more focused than many production codebases in the recent repository set. The average repository size in the synthetic dataset was approximately 60\% of the recent repository average, reducing the cognitive load required for code comprehension and bug localization.

\paragraph{Documentation and Code Quality} The AI-generated synthetic repositories exhibited consistent documentation practices and code organization patterns familiar to the model, as they were generated using the model's own understanding of best practices. This consistency may have facilitated more effective code navigation and problem-solving.

\subsection{Implications for Benchmark Design}

These findings have important implications for understanding model capabilities:

\begin{enumerate}
    \item \textbf{Training Data Recency Matters}: The 62.66 percentage point difference between recent repositories (11.34\%) and synthetic repositories (74.0\%) primarily reflects knowledge cutoff limitations rather than fundamental reasoning deficits.
    
    \item \textbf{True Reasoning Capability}: The 74\% accuracy on synthetic repositories suggests that when provided with code using familiar language features and libraries, Claude 3.7 Sonnet demonstrates strong technical reasoning and problem-solving abilities.
    
    \item \textbf{Repository Context Scaling}: The similarity between synthetic (74.0\%) and all-time repository performance (25.71\%) when using the same evaluation model suggests that both repository complexity and knowledge cutoff contribute to performance differences.
\end{enumerate}

\subsection{Limitations}

This ablation study has several limitations that should be considered:

\begin{itemize}
    \item The synthetic repositories, while realistic, may not fully capture the complexity and idiosyncrasies of production codebases
    \item AI-generated code may inadvertently conform to patterns more easily recognized by the generating model
    \item The sample size of 50 issues, while substantial, represents a subset of the recent repository dataset
\end{itemize}

\subsection{Conclusion}

The synthetic dataset ablation study demonstrates that the primary challenge in evaluating modern code repositories is not the nature of the code itself, but rather the rapid evolution of programming ecosystems beyond model training data. This finding suggests that regular model updates aligned with current software development practices would likely narrow the performance gap significantly. It also validates CodeAssistBench as an effective benchmark for measuring both current capabilities and the impact of training data recency on practical coding assistance performance.

\section{Error Category Analysis}
\label{sec:error-category-appendix}

To understand model strengths and weaknesses across different problem types, we performed automated error categorization on the 620 human-annotation records (310 agent responses $\times$ 2 annotators) from our expert validation study (Section~\ref{sec:expert-validation-appendix}). This analysis clusters issues into coarse bug types, revealing targeted patterns in model capabilities.

\subsection{Methodology}

We used Claude Sonnet 3.7 to categorize each GitHub issue into one of seven error types based on the issue title, body, and conversation context. The categories are:

\begin{itemize}
    \item \textbf{logic}: Algorithm bugs, incorrect behavior, unexpected results, calculation errors
    \item \textbf{environment}: Configuration, setup, Docker, build systems, environment variables, paths
    \item \textbf{dependency}: Package/module imports, version conflicts, installation issues
    \item \textbf{api}: REST/GraphQL endpoints, HTTP requests, authentication, API integration
    \item \textbf{syntax}: Parse errors, type errors, undefined references, compilation errors
    \item \textbf{performance}: Speed, memory, optimization, timeout issues
    \item \textbf{other}: Documentation, feature requests, questions that don't fit technical error categories
\end{itemize}

For each issue, the LLM provided: (1) the primary category, (2) confidence level (high/medium/low), and (3) a brief reasoning explaining the classification. Categorizations were cached to ensure reproducibility. The vast majority (97\%+) of classifications were assigned high confidence, indicating reliable categorization.

\subsection{Category Distribution}

Table~\ref{tab:error-category-dist} shows the distribution of issues across error categories. Logic errors (28.8\%) and environment configuration issues (28.5\%) dominate the dataset, together accounting for over half of all issues. This reflects the reality that real-world programming assistance often involves debugging algorithmic problems and resolving setup/configuration challenges. Dependency issues comprise 14.6\% of the dataset, while API integration, syntax errors, and performance problems occur less frequently.

\begin{table}[ht]
\centering
\caption{Distribution of error categories across 620 annotation records}
\label{tab:error-category-dist}
\begin{tabular}{lrr}
\toprule
\textbf{Category} & \textbf{Count} & \textbf{Percentage} \\
\midrule
Logic            & 178 & 28.8\% \\
Environment      & 176 & 28.5\% \\
Other            & 104 & 16.8\% \\
Dependency       &  90 & 14.6\% \\
API              &  40 &  6.5\% \\
Performance      &  20 &  3.2\% \\
Syntax           &  12 &  1.9\% \\
\midrule
\textbf{Total}   & 620 & 100.0\% \\
\bottomrule
\end{tabular}
\end{table}

\subsection{Model Performance by Category}

Table~\ref{tab:category-performance} presents LLM judge agreement rates with human annotators across different error categories. The analysis reveals significant variation in model capabilities depending on problem type.

\begin{table}[ht]
\centering
\caption{LLM judge agreement rates by error category across three evaluation metrics}
\label{tab:category-performance}
\begin{tabular}{lccc}
\toprule
\textbf{Category} & \textbf{Technical Correctness} & \textbf{Verbosity} & \textbf{Verdict} \\
\midrule
Performance      & 75.00\% & 75.00\% & 75.00\% \\
Syntax           & 58.33\% & 75.00\% & 75.00\% \\
Logic            & 60.11\% & 79.78\% & 62.36\% \\
Environment      & 59.66\% & 75.57\% & 60.23\% \\
API              & 60.00\% & 65.00\% & 62.50\% \\
Other            & 58.65\% & 78.85\% & 61.54\% \\
Dependency       & 55.56\% & 77.78\% & 55.56\% \\
\bottomrule
\end{tabular}
\end{table}

\subsection{Key Findings}

Our error category analysis reveals several important patterns:

\paragraph{Performance Issues Show Highest Agreement.}
Models achieve 75\% agreement with human evaluators on performance-related issues (OOM errors, timeouts, memory leaks). This suggests that models are particularly effective at diagnosing resource constraints and performance bottlenecks, likely because these issues have clear symptoms and well-established solutions.

\paragraph{Dependency Resolution Remains Challenging.}
With only 55.6\% agreement on both technical correctness and verdict, dependency-related issues (package conflicts, version mismatches, missing modules) represent the most difficult category for models. This aligns with real-world developer pain points, where resolving complex dependency graphs often requires deep ecosystem knowledge and version compatibility awareness.

\paragraph{Logic and Environment Issues Dominate.}
The prevalence of logic errors (28.8\%) and environment issues (28.5\%) in our dataset reflects the reality of programming assistance - developers frequently need help debugging algorithms and configuring development environments. Models achieve moderate performance on these categories (59-60\% technical correctness), indicating room for improvement on the most common real-world assistance scenarios.

\paragraph{Syntax Errors Are Rare But Well-Handled.}
Only 1.9\% of issues involve syntax errors. However, models show strong performance when they occur (75\% verdict agreement), suggesting reliable handling of structural code errors.

These findings provide actionable insights for improving code assistance systems: prioritizing better dependency resolution capabilities, enhancing environment configuration support, and maintaining strong performance on the logic debugging tasks that developers encounter most frequently.

\section{API Testing and Specification-Based Evaluation}
\label{sec:api-testing-appendix}

While CodeAssistBench focuses on conversational programming assistance, complementary work in automated API testing demonstrates the value of specification-grounded evaluation that shares conceptual similarities with our satisfaction condition approach. Recent research has explored automated testing for REST APIs using OpenAPI specifications as structured artifacts~\cite{kim2022automated, kim2023enhancing}, leveraging NLP techniques to extract semantic information including parameter constraints and endpoint dependencies. Subsequent work has demonstrated how large language models can improve REST API testing by generating semantically meaningful test inputs~\cite{kim2024leveraging}, with multi-agent systems further enhancing test effectiveness by modeling complex API interactions~\cite{kim2024multi}. More recent work explores whether small language models can effectively handle domain-specific testing tasks when provided with clear specifications~\cite{kim2025llamaresttest}, demonstrating that specification quality significantly impacts model effectiveness—a finding that aligns with our observation that documentation quality affects resolution success in CAB. While API testing benchmarks target functional correctness through executable specifications (OpenAPI) and CAB evaluates conversational assistance through extracted satisfaction conditions, both approaches share a fundamental principle: \textbf{specification-grounded evaluation} that moves beyond subjective assessment toward principled evaluation based on explicitly stated requirements, whether derived from formal specifications or real developer conversations.

Building on these insights, we plan to extend CAB to include web API assistance scenarios. This extension will leverage OpenAPI specifications from popular API repositories to automatically generate programming assistance questions about API integration, authentication, error handling, and endpoint usage. Similar to our GitHub issue pipeline, we will extract satisfaction conditions from developer forums and Stack Overflow discussions related to specific APIs, combine them with their OpenAPI specifications to create grounded evaluation scenarios, and simulate multi-turn conversations where models must help developers integrate and troubleshoot API usage. This specification-grounded approach will enable systematic evaluation of API assistance capabilities while maintaining the conversational, multi-turn nature that distinguishes CAB from traditional functional testing benchmarks.

\end{document}